\documentclass[10pt,conference]{IEEEtran}

\usepackage{cite}
\usepackage{amsmath,amssymb,amsfonts}
\usepackage{algorithmic}
\usepackage{graphicx}
\usepackage{textcomp}
\usepackage{xcolor}
\usepackage[hyphens]{url}
\usepackage{fancyhdr}
\usepackage{hyperref}
\usepackage{tikz}
\usepackage{booktabs}
\usepackage{tabularx}
\usepackage{multirow}
\usepackage{siunitx}
\usepackage{makecell}
\usepackage{array}
\usepackage{pifont}
\usepackage{threeparttable}

\newcommand{\hpcayear}{2027}
\newcommand{\hpcasubmissionnumber}{3592}
\title{You Only Charge Once 2.0 : A End-to-End Analog Computing-in-Memory Architecture with Reconfigurable Switched Capacitors}

\def\hpcacameraready{}

\newcommand\hpcaauthors{
    Zihao Xuan\IEEEauthorrefmark{1}\IEEEauthorrefmark{2},
    Yewen Li\IEEEauthorrefmark{1}\IEEEauthorrefmark{2},
    Jia Chen\IEEEauthorrefmark{3},
    Wei Xuan\IEEEauthorrefmark{4},
    Xiao Huo\IEEEauthorrefmark{2},
    Fengbin Tu\IEEEauthorrefmark{1}\IEEEauthorrefmark{2}
    }
\newcommand\hpcaaffiliation{ 
    \IEEEauthorrefmark{1}The Hong Kong University of Science and Technology, Hong Kong, China \\
    \IEEEauthorrefmark{2}AI Chip Center for Emerging Smart Systems (ACCESS), InnoHK, Hong Kong \\
    \IEEEauthorrefmark{3}Huazhong University of Science and Technology, Wuhan, China \\
    \IEEEauthorrefmark{4}Shenzhen University of Advanced Technology, Shenzhen, China
    }
\newcommand\hpcaemail{email: zihaoxuan@ust.hk, fengbintu@ust.hk}

\newcolumntype{C}{>{\centering\arraybackslash}X}

\newcommand{\sstab}{\rule{0pt}{8pt}\\[-1.8ex]}
\newcommand{\cmark}{\ding{51}}
\newcommand{\xmark}{\textcolor{red}{\ding{55}}}

\author{
  \ifdefined\hpcacameraready
    \IEEEauthorblockN{
      \hpcaauthors{}
    }
    \IEEEauthorblockA{
      \hpcaaffiliation{} \\
      \hpcaemail{}
    }
  \else
    \IEEEauthorblockN{\normalsize{HPCA \hpcayear{} Submission
      \textbf{\#\hpcasubmissionnumber{}}} \\
      \IEEEauthorblockA{
        Confidential Draft \\
        Do NOT Distribute!!
      }
    }
  \fi 
}

\begin{document}
\maketitle

\newcommand{\hpcaheight}{0mm}
\ifdefined\eaopen
\renewcommand{\hpcaheight}{12mm}
\fi

\begin{abstract}
Analog Computing-in-Memory (ACiM) accelerates deep neural networks by keeping weights inside memory arrays and executing dot products in the analog domain. However, modern ACiM accelerators are often limited by the "ADC wall": analog-to-digital converters consume a large fraction of energy and area, while bit-sliced execution repeatedly invokes these converters. Existing designs reduce this cost with low-resolution readout or time multiplexing, but they either lose output fidelity or introduce serialization overhead.

Charge-CIM addresses this bottleneck by using switched-capacitor charge redistribution as a uniﬁed computing and conversion substrate. The same capacitor fabric performs input conversion, analog MAC, weighted shift-and-add, and readout quantization, reducing both standalone converter overhead and intermediate ADC invocations.  A differential readout path further combines paired partial sums during ADC quantization, providing a highly compact and energy-efficient solution for array integration. With   dataflow architecture support, we evaluated Charge-CIM on a suite of DNN benchmarks, from CNNs to Transformer models, and experimental results show that Charge-CIM reduces ADC energy by 91.7\% under our evaluation setup and improves energy efficiency by 2.7$\times$ and throughput by 2.0$\times$ compared to the state-of-the-art charge-domain CIM accelerator.

\end{abstract}


\section{Introduction}


Computing in-Memory (CIM) has emerged as a promising paradigm to solve the "memory wall" problem by performing arithmetic operations directly within the memory array \cite{mutlu2022modern, kachris2025survey}. This approach eliminates the overhead of weight data movement and achieves efficient DNN acceleration \cite{sun2023survey}. Among various CIM implementations, Analog CIM (ACiM) is particularly promising \cite{sun2023analog, sun2023full}. By leveraging physical principles such as Kirchhoff's laws or charge conservation, ACiM can execute dot-product operations in the analog domain with low energy consumption, making it highly suitable for power-constrained scenarios \cite{Kang2020DeepIA, fick2022analog, zhao2024light}.

Despite their potential, ACiM architectures are severely hindered by analog-to-digital converter (ADC) overhead. Traditional ACiM designs rely heavily on power-hungry and area-intensive ADCs to interface between the analog computational core and the digital logic. Recent studies \cite{khwa201865nm, shafiee2016isaac, guo2024cambricon} show that ADCs can account for 50\% $\sim$ 70\% of the energy cost in high-throughput ACiM accelerators, significantly undermining the intrinsic efficiency benefits of ACiM.

The root cause of this overhead is the \textbf{ arithmetic slicing paradigm} \cite{xiao2023accuracy}. To map high-bit-width vector-matrix multiplication (VMMs)  onto low-resolution analog arrays, existing ACiM systems slice the computation along multiple dimensions: input bits are often processed over time, weight bits are mapped across columns, and large matrices are partitioned across arrays \cite{valavi201964}. Each slice produces an analog partial sum that must be digitized before digital shift-add or inter-array accumulation. Although slicing lowers the resolution required by each ADC, it increases the number of ADC invocations and exposes every partial result to quantization (e.g., 64 conversions with 1-bit/slice input and weight for 8-bit MAC).   Recent works such as RAELLA \cite{andrulis2023raella} and Cambricon-CIM \cite{guo2026cambricon} attempt to mitigate these issues via speculative recovery or non-binary encoding and lower ADC resolutions, they remain fundamentally constrained by sliced arithmetic. Consequently, they also suffer from multi-slice latency, frequent conversion overhead, and additional digital circuit costs (e.g., over 16\% system power overhead for encoding in Cambricon-CIM \cite{guo2026cambricon}).

In this paper, we answer a fundamental question to overcome these limitations: \textbf{Can we enable fully analog multi-bit computation to avoid arithmetic slicing and reduce ADC overhead?} The key idea is to reuse the switched-capacitor networks to create a unified analog compute and and conversion substrate.

To this end, we propose Charge-CIM,  a charge-domain ACiM architecture for fully analog multi-bit computation with reduced ADC overhead. Our main contributions are:

\sstab (1) \textbf{Fully analog computation alleviate slice-induced ADC overhead.} We introduce a novel mechanism for fully analog multi-bit computation via sequential charge sharing in a reconfigurable switched-capacitor network. This approach unifies digital-to-analog converter (DAC), multiplication, accumulation, and weighted summation in the analog domain, eliminating bit-slicing and reducing ADC conversion frequency.

\sstab (2) \textbf{Embedded ADC reduce per-converter cost.}  We reuse the switched capacitors inside the array as the Capacitive-DAC (C-DAC) , eliminating the most area- and energy-intensive part of a standalone ADC. This reduces per-channel ADC overhead while preserving parallel output readout.

\sstab (3) \textbf{Differential macro with in-ADC accumulation to improve ADC precision.} We combine a differential macro with inverse encoding so that analog partial sums are added during ADC conversion. This reduces conversion steps and improves noise immunity through differential readout.

\sstab (4) \textbf{Full-stack architectural integration.} We build a hierarchical Charge-CIM architecture across macro, In-situ Multiply-Accumulator (IMA), tile, and chip levels, with special function units (SFU) integration.  This enables flexible layer-fusion dataflows, including GEMM-GEMM, GEMM-non-GEMM, and non-GEMM operator fusion.

Finally, we evaluate Charge-CIM performance across nine DNN benchmarks. Simulation results demonstrate that, compared to state-of-the-art (SOTA) baselines RAELLA \cite{andrulis2023raella} and Cambricon-CIM \cite{guo2026cambricon}, Charge-CIM achieves an average energy efficiency improvement of $6.1\times$ and $2.7\times$, and $7.3\times$ and $2.0\times$ higher throughput, respectively, while reaching an overall compute density of $1.4~\text{TOPS/mm}^2$.

\section{Background and Motivation}
\label{sec:background}

\subsection{Deep Neural Network Workloads}

Deep Neural Networks (DNNs), including Convolutional Neural Networks (CNNs) and Transformers, are dominated by Vector-Matrix Multiplication (VMM) over large parameter sets~\cite{krizhevsky2012imagenet,vaswani2017attention,he2016deep}. As model size increases, efficient inference requires high-throughput parallel computation~\cite{sevilla2022compute}. INT8 quantization is widely adopted to reduce storage and data movement while maintaining model accuracy~\cite{zhu2020towards,xiao2023smoothquant}. Mapping INT8 multiply-accumulate operations directly onto Analog Computing-in-Memory (ACiM) arrays further improves energy efficiency by reducing memory access overhead~\cite{sun2025model}.

\subsection{Charge-Domain ACiM Basics}

\begin{figure}[htbp]
    \centering
    \hspace{-0.045\linewidth}\includegraphics[width=1.05\linewidth]{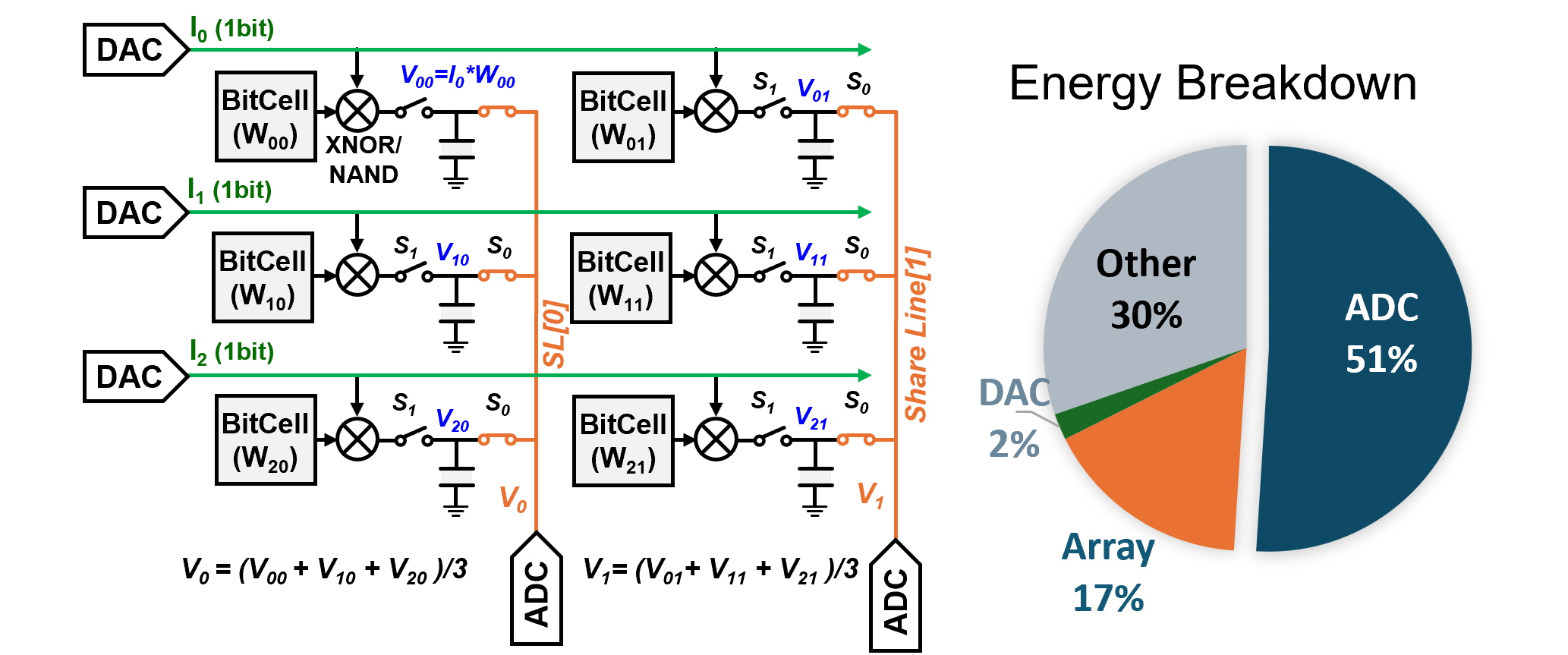}
    \caption{(Left) Conceptual diagram of a charge-domain Compute-in-Memory (CiM) array. Inputs ($I_i$) and weights ($W_{ij}$) are multiplied to generate local voltages ($V_{ij}$), which are then averaged across the column ($V_j = \sum V_{ij} / N$) using charge-sharing capacitors before being converted to digital outputs by ADCs. (Right) Energy breakdown in a conventional charge-domain SRAM-based CIM design \cite{guo2026cambricon}, where ADCs account for over 51\% energy cost.}
    \label{fig:basic_charge_acim}
\end{figure}

Unlike current-domain ACiM~\cite{si2021local,9960777}, which accumulates transistor-driven bitline currents, charge-domain ACiM performs computation through switched-capacitor charge redistribution. Since the result is primarily determined by capacitor ratios, charge-domain computing provides high linearity and reduced sensitivity to PVT variations~\cite{chen2021cap}. Figure~\ref{fig:basic_charge_acim} left shows a typical charge-domain ACiM array consisting of input DACs, bitcells, multiplication units, switched-capacitor networks, and output ADCs. Each capacitor samples the multiplication result
$V_{i,j}=I_i \cdot W_{i,j}$ and stores the corresponding charge
$Q=C_u \cdot V_{i,j}$. When the capacitors in the same column are connected, charge redistribution produces the column voltage
$V_j=\sum_i V_{i,j}/N$, which represents the averaged partial sum. The resulting voltage is then digitized by an ADC.

\subsection{Slicing-Induced ADC Overhead}

\begin{figure*}
    \centering
    \includegraphics[width=0.9\linewidth]{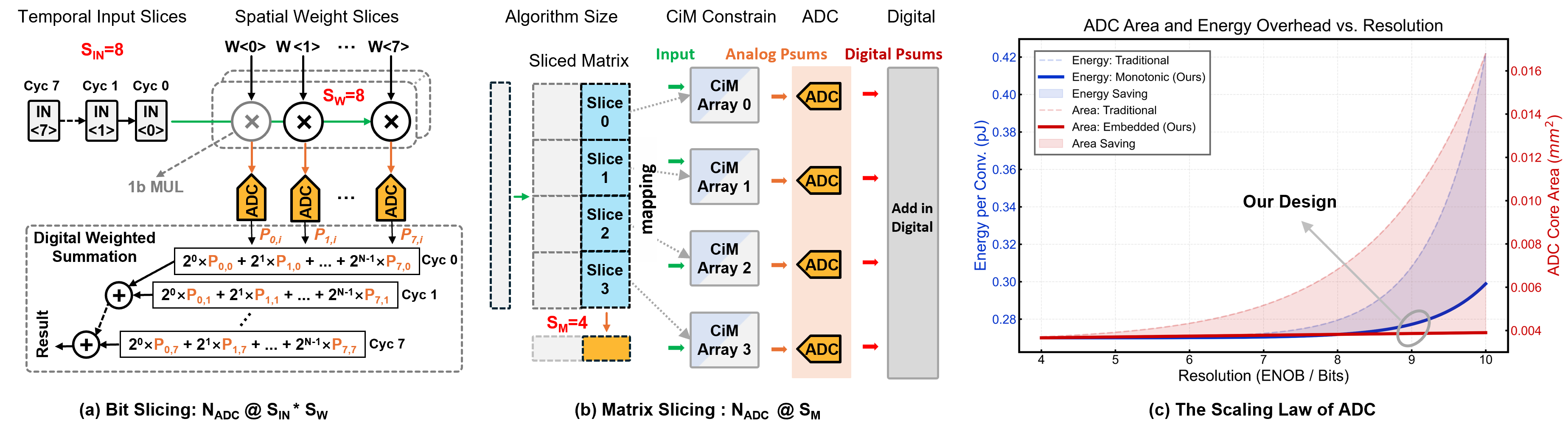}
    \caption{Overview of CiM mapping strategies and ADC overhead. (a) Bit slicing for MAC operations. (b) Matrix slicing for large-scale algorithms. (c) Energy and area scaling laws of ADCs versus resolution, comparing traditional designs with the proposed approach.}
    \label{fig:slice_adc}
\end{figure*}

According to RAELLA~\cite{andrulis2023raella}, the primary ADC overhead arises from frequent conversions caused by arithmetic slicing. As shown in Figure~\ref{fig:slice_adc}, existing approaches mainly include bit-slicing and matrix-slicing.

\textbf{In bit-slicing}, inputs and weights are divided into multiple slices. Input slices are processed temporally across multiple cycles, while weight slices are mapped spatially to different array columns. Each analog result is digitized and reconstructed using a shift-adder. Therefore, the number of ADC conversions scales with the product of input and weight slices, i.e., $N_{\mathrm{ADC}} \propto S_{\mathrm{IN}} \cdot S_{\mathrm{W}}$. Figure~\ref{fig:slice_adc}(a) shows an extreme case in which an 8-bit input is processed over eight cycles and an 8-bit weight is distributed across eight columns, requiring 64 conversions per output dot product. Although higher-resolution DACs can reduce input slicing, they introduce additional area and energy overhead.

\textbf{In matrix-slicing}, large weight matrices are partitioned into $S_{\mathrm{M}}$ blocks due to the limited array size, as shown in Figure~\ref{fig:slice_adc}(b). Each block independently digitizes its partial sum before digital accumulation, causing the ADC overhead to scale linearly with the number of matrix slices, i.e., $N_{\mathrm{ADC}} \propto S_{\mathrm{M}}$. Overall, the conversion frequency scales as $N_{\mathrm{ADC}} \propto S_{\mathrm{IN}} \cdot S_{\mathrm{W}} \cdot S_{\mathrm{M}}$. As illustrated in Figure~\ref{fig:slice_adc}(c), because ADC area and energy increase exponentially with resolution, i.e., $2^N$, slicing-induced conversions create an ``ADC wall'' that offsets the efficiency benefits of analog computation.

\subsection{Limitations of Current Works}

Prior ACiM optimizations mainly follow two directions: circuit primitives and architecture-level mitigation. Circuit-level solutions like C2C-CIM~\cite{wang2023charge} adopts a C-2C ladder for 8-bit weighted MACs; PICO-RAM~\cite{chen2024pico} and YOCO~\cite{xuan2025yoco} reuse local capacitors for analog shift-and-add, and CR-CIM~\cite{yoshioka2024818} explores capacitor reconfiguration to reduce ADC overhead and improve area efﬁciency. However, these designs usually optimize only part of the analog datapath.  Supporting high-precision multi-bit computation still requires standalone ADC readout, bit-slicing, digital reconstruction, or extra compensation logic, with ADC-related operations still accounting for up to approximately 51\% of the total energy (see Figure~\ref{fig:basic_charge_acim}  right).  

Architecture-level designs like  Cambricon-CIM~\cite{guo2026cambricon} and RAELLA~\cite{andrulis2023raella} lower ADC resolution with new encoding schemes, while TIMELY~\cite{li2020timely} distributes ADC cost through local integration and analog buffering. These methods reduce conversion cost, but introduce digital reconstruction overhead (occupy 40\%-50\% of the total digital logic), compensation logic, or signal-margin loss.  Table~\ref{tab:cim_architectures}  summarizes this gap across circuit features and architecture support.

\subsection{Motivation}

These limitations motivate a holistic analog computing paradigm that unifies the entire VMM operation before a single ADC conversion. Our approach is driven by two key insights: First, multi-dimensional charge redistribution enables slice-free multi-bit arithmetic, while the in-situ reuse of array capacitors as ADC components minimizes interface overhead. Second, by maintaining full-precision computation in the analog domain, we preserve numerical fidelity and avoid the premature information loss typical of frequent intermediate digitizations. This strategy mirrors the digital "high-precision accumulation followed by quantization" framework, ensuring INT8-level accuracy while maximizing energy efficiency.

\section{Charge-CIM Design}

\subsection{Principle of Charge-CIM Array}

\begin{figure}[t]
    \centering
    \includegraphics[width=0.8\linewidth]{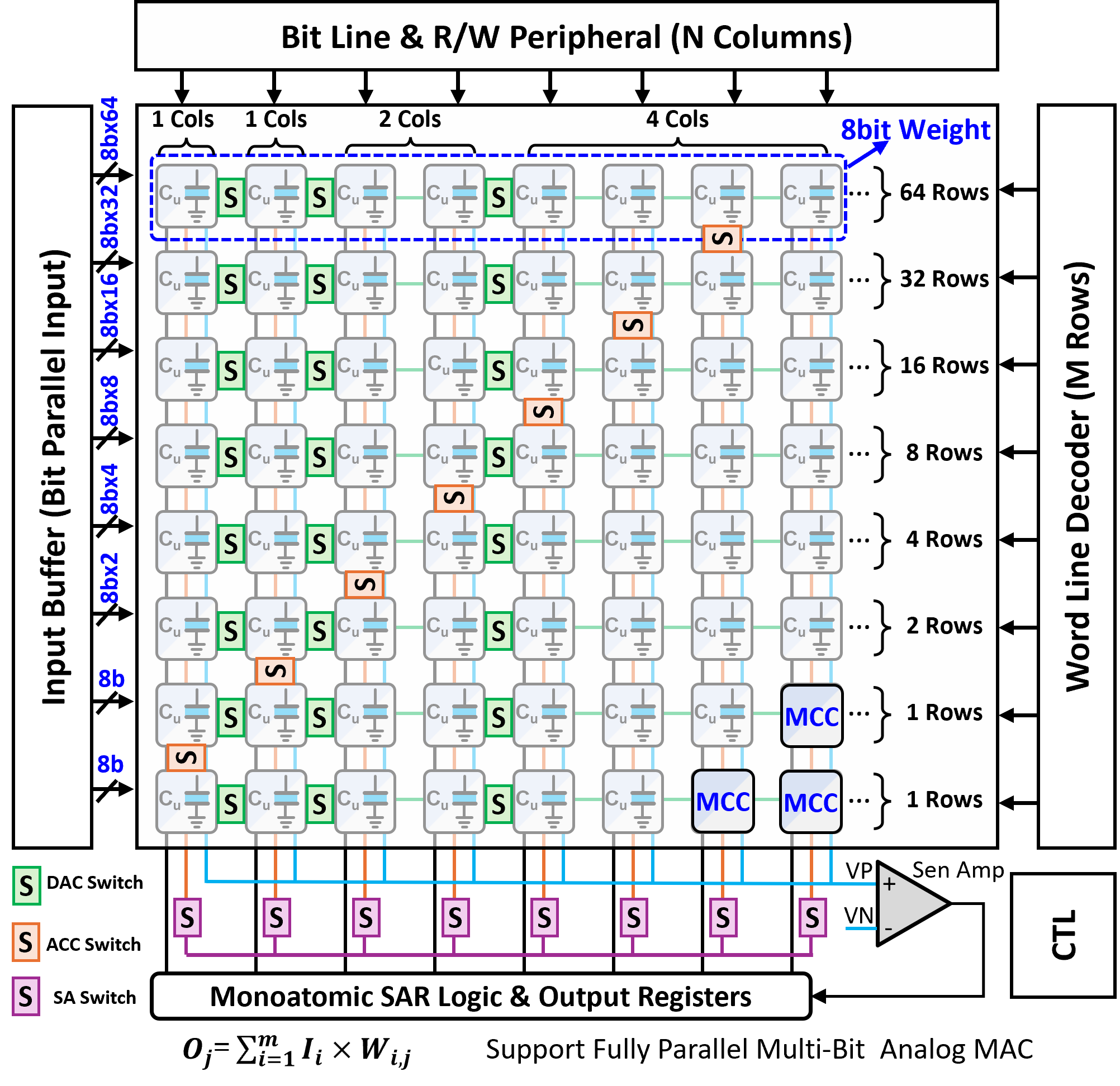}
    \caption{The proposed CiM macro architecture, illustrating the integration of the computing cell array, bit-parallel input buffer, and peripheral circuits, including the monotonic SAR logic.}
    \label{fig:array}
    \vspace{-1mm}
\end{figure}

\begin{figure}[t]
    \centering
    \includegraphics[width=0.9\linewidth]{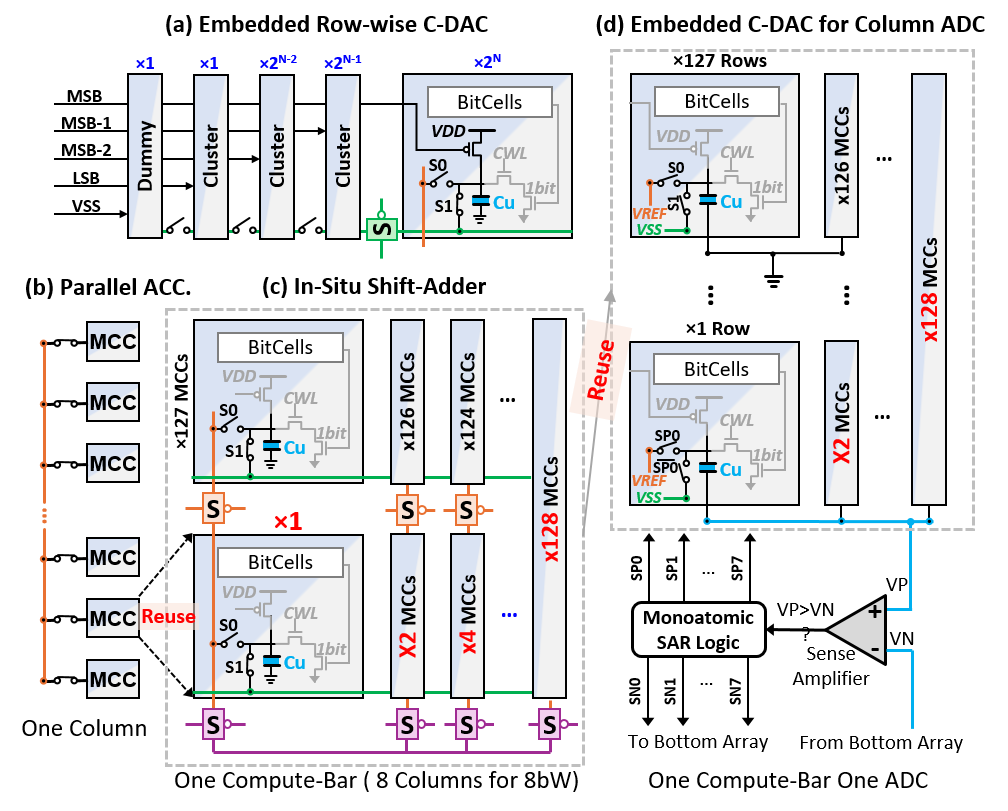}
    \caption{Details of the core computing circuits, showing the reconfigurable C-DAC structures for row/column operations and the SAR ADC logic.}
    \label{fig:reconfigure_cap}
        \vspace{-1mm}
\end{figure}

Conventional charge-domain ACiM uses unit capacitors $C_u$ for column-wise accumulation, while data conversion and weighted multiplication rely on peripheral circuits, limiting parallelism and energy efficiency. To address this issue, we propose a unified architecture that maximizes capacitor reuse. As shown in Figure~\ref{fig:array}, each Charge-CIM macro contains an array of memory-and-compute cells (MCCs). Each MCC integrates weight-storage bitcells and a unit switched capacitor, with multi-bit weights mapped across adjacent MCCs along each row. A lightweight distributed switch network, including $S_{DAC}$, $S_{ACC}$, and $S_{SA}$, dynamically reconfigures these capacitors for different operations, as detailed below:


\sstab (1) \underline{\textit{Row-wise (Figure~\ref{fig:reconfigure_cap}a).}} $2^N$ MCCs are grouped via $N$ $S_{DAC}$ switches in a binary ratio ($1:1:2:\dots:2^{N-1}$), functioning as an $N$-bit embedded DAC for input conversion.

\sstab (2) \underline{\textit{Column-wise (Figure~\ref{fig:reconfigure_cap}b).}} Charge sharing enables parallel accumulation and averaging across the column.

\sstab (3) \underline{\textit{Compute-Bar (Figure~\ref{fig:reconfigure_cap}c).}} $M$ columns are organized as a compute-bar (CB) for an $M$-bit weight vector. In one CB, $S_{SA}$ switches configure the shared capacitance ratio between columns to $1:2:\dots:2^{M-1}$, realizing in-situ shift-and-add.

\begin{table}[htbp]
\centering
\resizebox{1\columnwidth}{!}{
\begin{threeparttable}
\caption{Comparison of features among different ACiM architectures.\tnote{\dag}}
\label{tab:cim_architectures}
\begin{tabular}{|l|c|c|c|c|c|}
\hline
& \textbf{\makecell{Embedded \\ DAC}} & \textbf{\makecell{In-situ \\ SA}} & \textbf{\makecell{Embedded \\ ADC}} & \textbf{\makecell{In-ADC \\ Psum Add}} & \textbf{\makecell{Architecture \\ Support}} \\
\hline
C2C-CIM \cite{wang2023charge} & \xmark & \cmark & \xmark & \xmark & \xmark \\ \hline
CR-CIM \cite{yoshioka2024818} & \xmark & \xmark & \cmark\tnote{\ddag} &  \xmark & \xmark \\ \hline
PICO-RAM \cite{chen2024pico} & \cmark & \cmark & \xmark & \xmark & \xmark \\ \hline
Cambricon-CIM \cite{guo2026cambricon} & \xmark & \cmark\tnote{\dag} & \xmark & \xmark & \cmark \\ \hline
RAELLA \cite{andrulis2023raella} & \xmark & \cmark & \xmark &  \xmark & \cmark\tnote{*} \\\hline
TIMELY \cite{li2020timely}& \xmark& \xmark& \xmark& \xmark&\cmark\\\hline 
YOCO \cite{xuan2025yoco} & \cmark & \cmark & \xmark & \xmark & \cmark \\ \hline
\textbf{Charge-CIM} & \textbf{\cmark} & \textbf{\cmark} & \textbf{\cmark} & \textbf{\cmark} & \textbf{\cmark} \\ \hline
\end{tabular}

\begin{tablenotes}[flushleft]
\small
\item[\dag] Cambricon-CIM adopts a C-ladder capacitor network for multi-bit MAC.
\item[\ddag] CR-CIM adopts a unit capacitor for 1-bit MAC.
\item[*] RAELLA performs analog MAC in memristors.
\end{tablenotes}
\end{threeparttable}
}
\end{table}

\sstab (4) \underline{\textit{Quantization (Figure~\ref{fig:reconfigure_cap}d).}} The same capacitor array in feature 3 serves as the C-DAC for an embedded SAR-ADC, enabling in-situ quantization.

\begin{figure}
    \centering
    \includegraphics[width=0.9\linewidth]{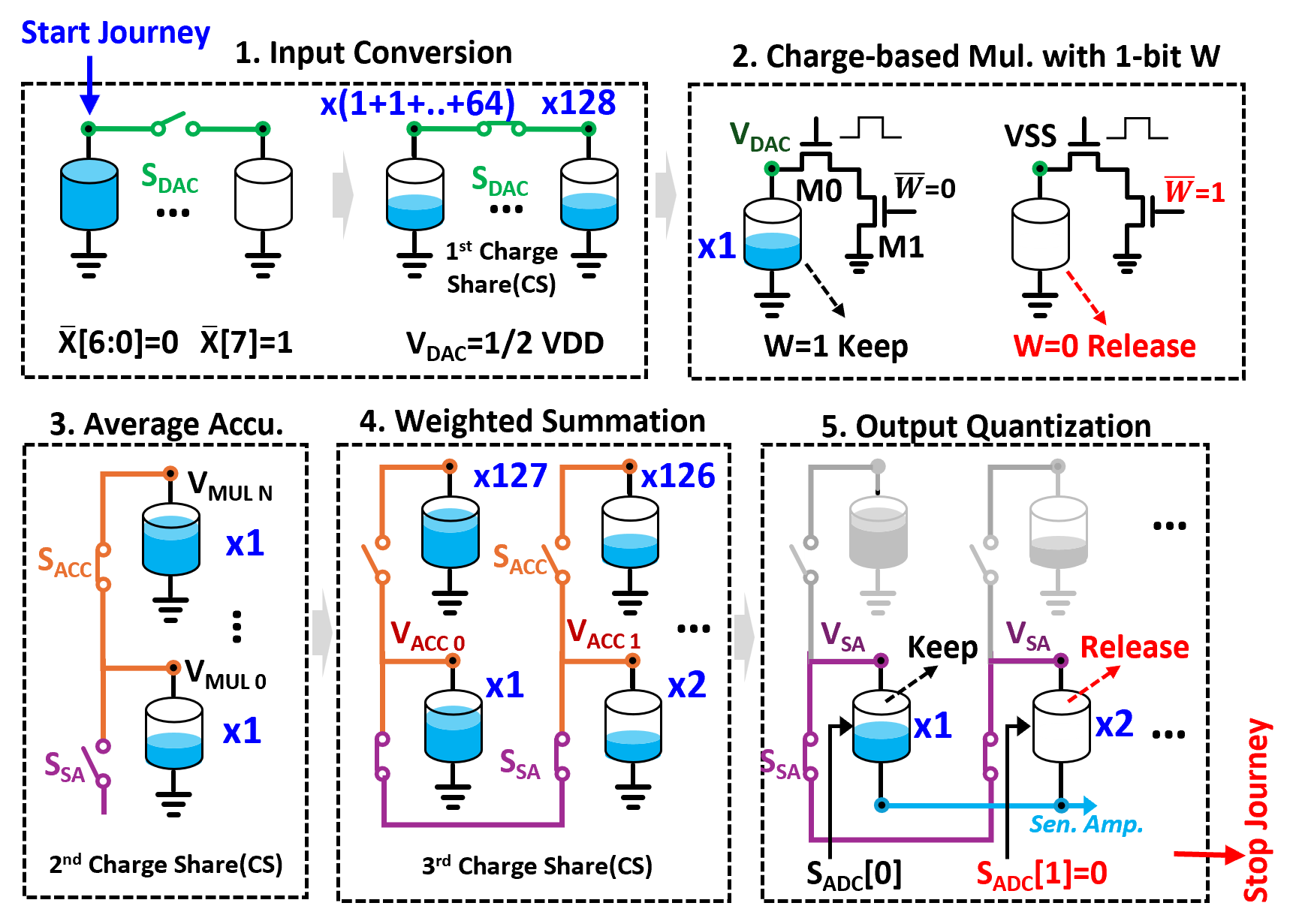}
    \caption{Six-phase operation of the proposed charge-based multi-bit VMM and output quantization.}
    \label{fig:life_charge}
\end{figure}

Figure~\ref{fig:life_charge} provides a step-by-step illustration of the charge-based MAC operation, which consists of six phases:

\textbf{Phase 1: Input Charging.} The $S_{DAC}$ switches are initially open. Depending on the input code, binary-sized capacitor groups are selectively charged to $V_{DD}$ through a PMOS transistor or left uncharged (see Figure~\ref{fig:life_charge}(1)).

\textbf{Phase 2: Embedded DAC.} The $S_{DAC}$ switches are then closed, triggering row-wise charge sharing and producing a uniform analog input voltage without a standalone DAC. For example, the 8-bit input $X[7:0]=\text{``01111111''}$ produces $127/256\,V_{DD}$, which is approximately $V_{DD}/2$ (see Figure~\ref{fig:life_charge}(2)). The input voltage for the $i$-th row is:
\begin{equation}
    V_{i}^{in} = \frac{1}{C_{tot}} \sum_{n=0}^{N-1} C_n \cdot V_n = V_{DD} \cdot \left( \frac{1}{2^N} \sum_{n=0}^{N-1} 2^n \cdot X_i[n] \right)
    \label{eq:isdac}
\end{equation}
where $X_i[n] \in \{0, 1\}$ is the $n$-th bit of the input; $V_n = V_{\mathrm{DD}} \cdot X_i[n]$; $C_n = 2^n \cdot C_u$ represents the total capacitance of the $n$-th group; and $C_{\mathrm{tot}} = 2^N \cdot C_u$ is the total row capacitance. This embedded scheme is  scalable and  suited for a dense array.

\textbf{Phase 3: Multiplication with a 1-bit Weight.} In each MCC, analog multiplication is performed by a compact NAND-like structure consisting of NMOS $M_0$ and $M_1$. While $M_0$ is controlled by the compute word line (CWL), the state of $M_1$ is determined by the stored 1-bit weight ($W$). During evaluation ($S_0$ and $S_1$ open), the MCC performs conditional charge retention: if $W=1$, $M_1$ is cut off to retain the charge; otherwise, the capacitor discharges. This mechanism multiplies the analog input voltage $V_i^{in}$ by the binary weight $W_{ij}[m]$. The resulting voltage on the local capacitor, representing the partial product, is:
\begin{equation}
    V_{ij}^{mul}[m] = V_i^{in} \cdot W_{ij}[m]
    \label{eq:mul}
\end{equation}
where $V_{i}^{in}$ is the $i$-th row input voltage and $W_{ij}[m]$ is the $m$-th weight bit. This approach stores intermediate products as charges for the subsequent accumulation phase (see Figure~\ref{fig:life_charge}(3)).

\textbf{Phase 4: Parallel Accumulation.} After multiplication, switches $S_0$ and $S_{ACC}$ are closed (Figure~\ref{fig:life_charge}(4)), while $S_{SA}$ remains open. This operation connects all MCCs within the same column, allowing their capacitors to share charge. Consequently, the shared-node voltage converges to the average of the multiplication voltages in that column:
\begin{equation}
    V_j^{acc}[m] = \frac{1}{P} \sum_{i=0}^{P-1} V_{ij}^{mul}[m]
    \label{eq:acc}
\end{equation}
\noindent where $V_{ij}^{mul}[m]$ denotes the multiplication voltage associated with the $m$-th weight bit at row $i$ and column $j$, as defined in Equation~\eqref{eq:isdac}, and $P$ is the number of rows in the MCC array. To support an $M$-bit weight, $P$ must satisfy $P>2^{M-1}$.

\textbf{Phase 5: Weighted Summation.}
After Phase 4, each bit column produces an analog MAC result corresponding to one weight bit. To obtain a multi-bit-weight VMM result, the $S_{ACC}$ switches are opened and the $S_{SA}$ switches are closed, as shown in Figure~\ref{fig:life_charge}(5). The unit capacitors across bit columns are grouped according to the binary-weighted ratio $1:2:4:\dots:2^{M-1}$. The shared voltage represents the bit-weighted sum of the column voltages:
\begin{equation}
    V_j^{out} = \frac{1}{2^M - 1} \sum_{m=0}^{M-1} 2^m \cdot V_{j}^{acc}[m]
    \label{eq:sa}
\end{equation}
\noindent where $M$ is the bit width of the weight and $V_j^{acc}[m]$ is the accumulated voltage of the $m$-th bit column. Substituting Equations~\eqref{eq:isdac}, \eqref{eq:mul}, and \eqref{eq:acc} into Equation~\eqref{eq:sa} gives the end-to-end multi-bit VMM expression:
\begin{equation}
    V_j^{out} = \frac{ \sum_{m=0}^{M-1} \sum_{i=0}^{P-1} \sum_{n=0}^{N-1} \left( 2^n X_i[n] \cdot 2^m W_{ij}[m] \right) }{ (2^M - 1) \cdot P \cdot 2^N } V_{DD}
    \label{eq:vmm_total}
\end{equation}
\noindent where $X_i$ is the $i$-th row input with $N$-bit precision, and $W_{ij}$ is the weight at row $i$ and column $j$ with $M$-bit precision.

\textbf{Phase 6: In-situ ADC Quantization.}
In the final phase, the hierarchical capacitor groups established in Phase 5 are repurposed as the binary-weighted capacitor array (C-DAC) of an embedded successive-approximation-register (SAR) ADC. As illustrated in Figure~\ref{fig:life_charge}(6), the bottom plates of these groups are routed to the input of a sense amplifier. Under the control of monotonic SAR logic, the array performs a successive-approximation search and quantizes the multi-bit analog voltage in situ. The quantization is:
\begin{equation}
    D_{j}^{out} = \text{quant}\left( V_j^{out} \right) = \left\lfloor \frac{V_j^{out}}{V_{LSB}} \right\rfloor = \left\lfloor \frac{V_j^{out}}{V_{REF} / 2^K} \right\rfloor
    \label{eq:quant}
\end{equation}
\noindent where $D_{j}^{out}$ is the $K$-bit quantized digital output for the $j$-th column, $V_{LSB}$ is the SAR-ADC voltage resolution, and $V_{REF}$ is the comparison reference voltage. This reuse eliminates the C-DAC area of a standalone ADC.

\textbf{The Journey of Charge.}
In summary, this six-phase operation performs multi-bit VMM and output conversion along one charge-processing path. Charge is injected from the supply in Phase 1; during Phases 2--5, it undergoes row- and column-wise redistribution for input conversion, multiplication, accumulation, and bit-weighted summation. In Phase 6, monotonic SAR logic discharges the capacitor groups to quantize the result in situ. This ``one-charge, multi-compute'' path avoids recharging between intermediate arithmetic stages.

\subsection{Differential ADC and Monotonic SAR Logic}
\begin{figure}[t]
    \centering
    \includegraphics[width=0.9\linewidth]{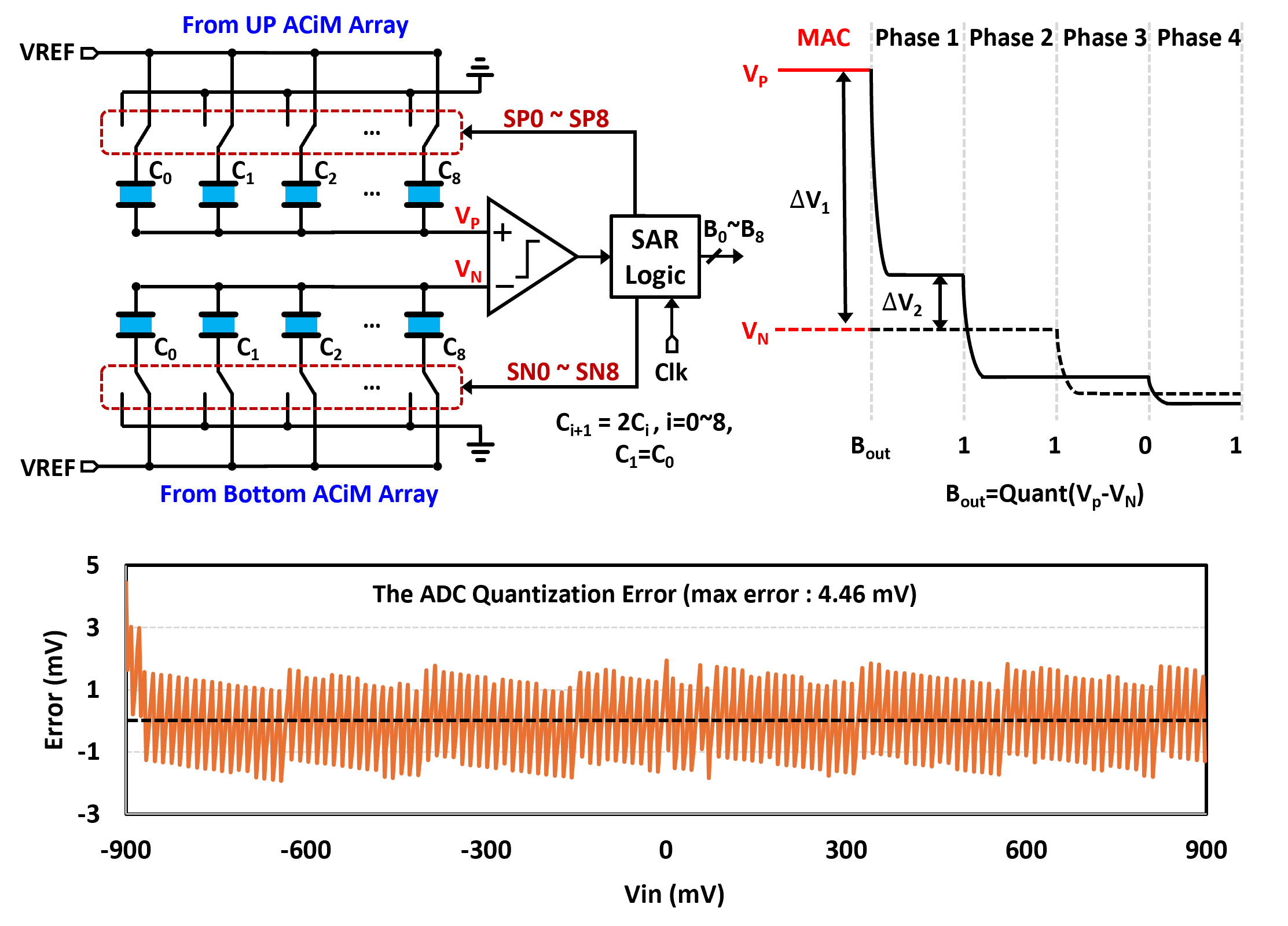}
    \caption{The proposed SAR ADC design, featuring the capacitor array schematic, conversion process, and quantization error analysis.}
    \label{fig:mono_adc}
\end{figure}

To improve noise immunity and precision, Charge-CIM employs a fully differential Embedded SAR ADC architecture.   As illustrated in Figure~\ref{fig:mono_adc}, outputs from two symmetric arrays are sampled at the positive  ($V_P$) and negative ($V_N$) comparator inputs through \textit{bottom-plate sampling}. A monotonic successive approximation scheme \cite{liu201010} selectively discharges capacitor groups to ground according to the comparator decisions, avoiding redundant charging from $V_{REF}$ and reducing both switching energy and capacitance. 

The differential ADC naturally performs subtraction, which is converted into the required neural-network addition through the \textit{inverse coding} scheme described in Section~\ref{subsec:diff_adc_coding}. By synergizing the differential structure with the monotonic algorithm, our architecture maintains high linearity across the entire analog processing chain. Simulation results demonstrate a peak quantization error of only 4.46 mV across the full input range, and shows more than 8-bit effective number of bits (ENOB).

\subsection{Reconfigurable Bit-width and Precision}

Charge-CIM supports configurable input and weight precision without hardware modification. As shown in Table~\ref{tab:reconfigurable_bitwidth}, input precision is configured by mapping digital bits to the embedded DAC capacitor groups. Taking a 4-bit configuration as an example, the standard mode maps ${B_3, B_2, B_1, B_0}$ to binary-weighted capacitors for full precision. Lower-precision modes replicate input bits across capacitor groups; for example, the 2-bit parallel mode uses ${B_1, B_0, B_1, B_0}$ to improve robustness or throughput. Charge-CIM also supports non-binary bit shuffling for customized value representations and base reformation.

\begin{figure*}
    \centering
    \includegraphics[width=0.95\linewidth]{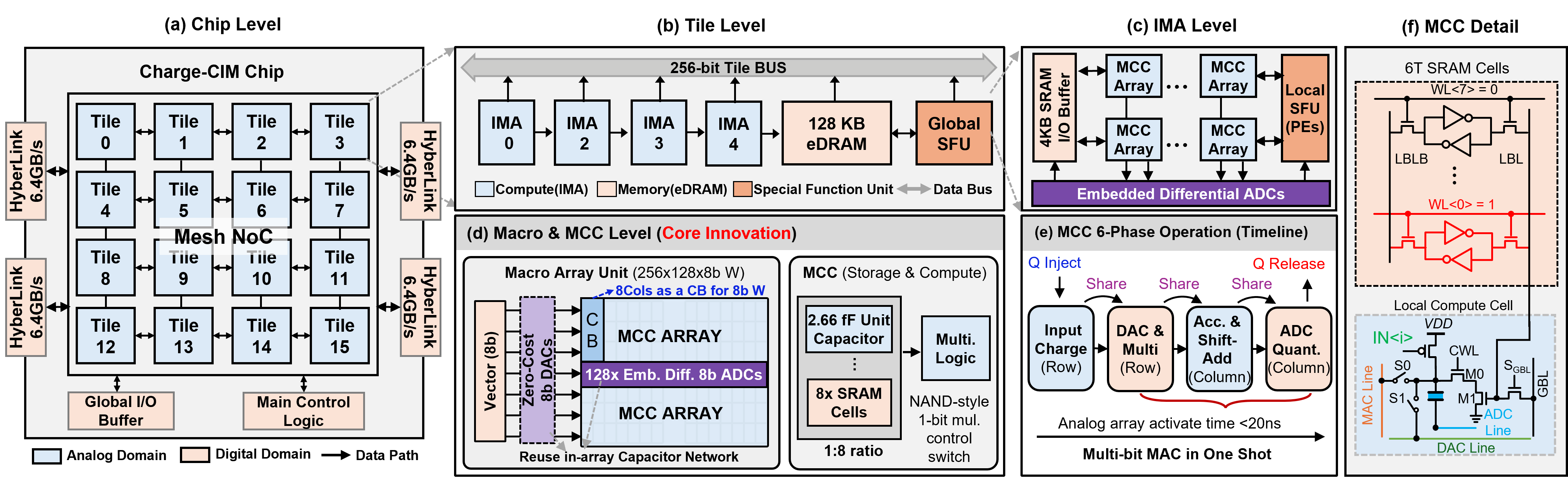}
    \caption{Hierarchical architecture of the proposed Charge-CIM accelerator: (a) chip, (b) tile, (c) IMA, (d) macro and MCC, (e) six-phase MCC operation, and (f) MCC circuit.}
    \label{fig:arch_overall}
\end{figure*}

\begin{table}[!t]
    \centering
    \caption{Configurable bit-widths and precision.}
    \label{tab:reconfigurable_bitwidth}
    \renewcommand{\arraystretch}{1.2}
    \setlength{\tabcolsep}{1.5pt}
    \fontsize{8}{8}\selectfont
    \begin{tabular}{@{} l c l c @{}}
        \toprule
        \textbf{Precision} &
        \textbf{Mode} &
        \textbf{Value Representation ($V_{\mathrm{in}}$)} &
        \textbf{Target} \\
        \midrule
        4-bit (Std) & $\{B_{3,2,1,0}\}$ & $B_3 2^3 + B_2 2^2 + B_1 2^1+ B_0 2^0$ & High Prec. \\
        2-bit (Par.) & $\{B_{1,0,1,0}\}$ & $B_1 (2^3+2^1) + B_0 (2^2+2^0)$ & Balanced \\
        1-bit (Binary) & $\{B_{0,0,0,0}\}$ & $ B_0 (2^3+2^2+2^1+2^0)$ & High Thrpt. \\
        \midrule
        \textbf{Non-Uniform} & $\{B_{3,0,2,1}\}$ & $B_3 2^3 + B_0 2^2 + B_2 2^1+ B_1 2^0$ & Base Reform. \\
        \bottomrule
    \end{tabular}
\end{table}

On the weight side, the precision is configured through the distributed $S_{SA}$ switch network. Selectively enabling the switches changes the number of columns involved in the final charge-sharing operation, thereby supporting weight precisions from 1-bit to $M$-bit. The independent configurability of input and weight precision allows Charge-CIM to balance accuracy, throughput, and energy efficiency across different workloads.

\subsection{Signed Arithmetic with Offset Encoding}
\label{subsec:signed_arithmetic}

To support signed arithmetic, Charge-CIM applies \textit{offset encoding} to both inputs and weights, where an N-bit signed value is represented by its unsigned code minus $2^{N-1}$. The signed dot product is therefore reformulated as:
\begin{equation}
\begin{aligned}
    O &= \sum_{i=0}^{L-1} (I_iW_i)^{sign} = \sum_{i=0}^{L-1}(I_i^{u} - 2^{N-1})(W_i^{u} - 2^{N-1}) \\
      &= \sum_{i=0}^{L-1} (I_i W_i)^{u}
      - 2^{N-1} \sum_{i=0}^{L-1} W_i^{u}
      - 2^{N-1} \sum_{i=0}^{L-1} I_i^{u}
      + 2^{2N-2}L
\end{aligned}
\label{eq:signed_arith}
\end{equation}
where $N$ is the bit width of both inputs and weights, and $L$ is the vector length. The four terms are implemented as follows: (1) \underline{\textit{Term 1 (Unsigned MAC).}} The Charge-CIM array directly computes the unsigned dot product. (2) \underline{\textit{Term 2 (Weight Compensation).}} Because the weights are stationary, their sum is precomputed during weight loading and stored in local registers. (3) \underline{\textit{Term 3 (Activation Compensation).}} A peripheral digital adder tree computes the input sum online. (4) \underline{\textit{Term 4 (Constant Bias).}} The constant term is merged into the post-processing bias and requires no runtime computation. This decomposition supports signed arithmetic with limited additional hardware.


\section{Architecture and Dataflow}
\label{sec:architecture}

\begin{figure}
    \centering
    \includegraphics[width=0.9\linewidth]{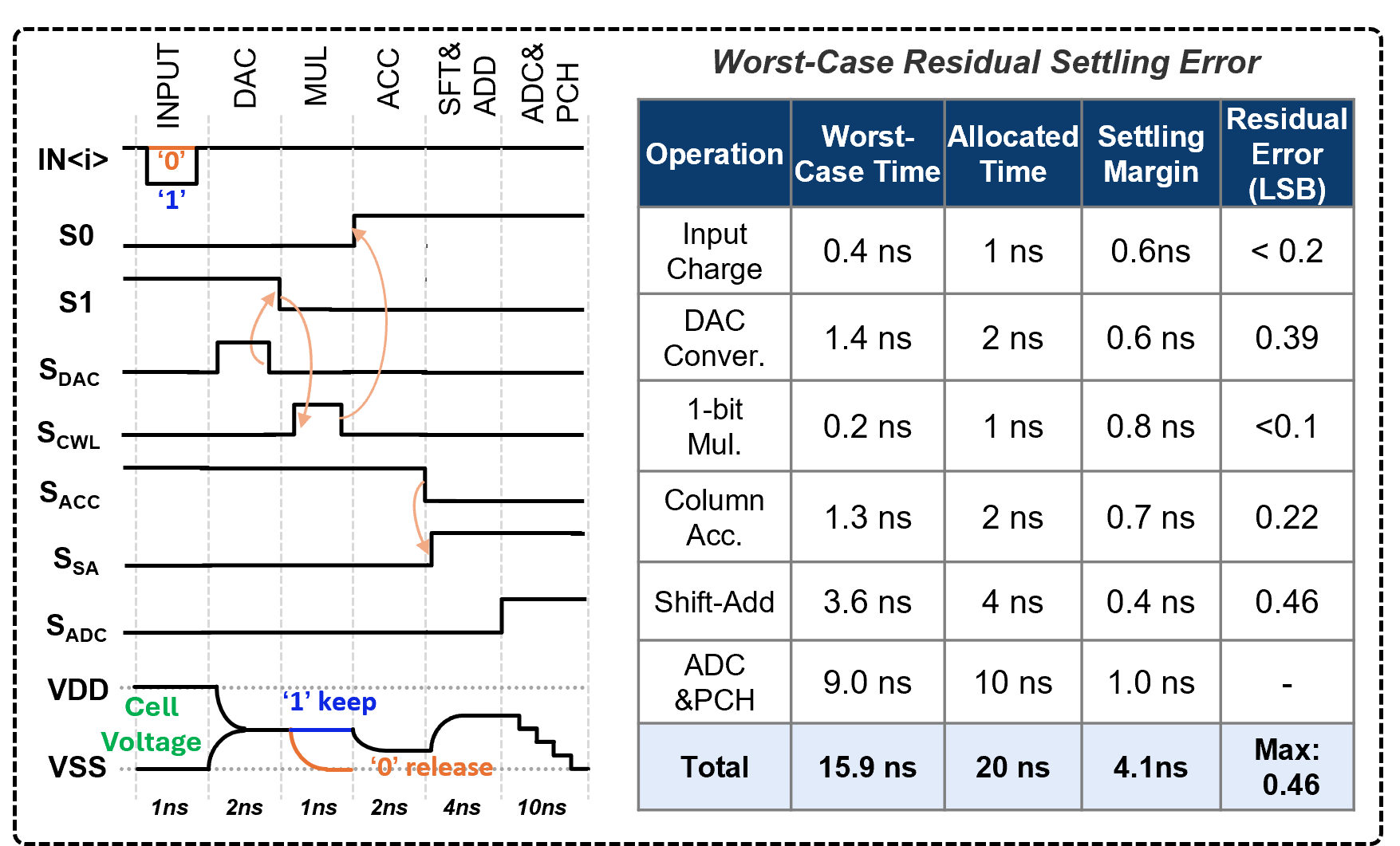}
    \caption{MCC operation wave and residual settling error.}
    \label{fig:operation_wave}
\end{figure}

\subsection{Hierarchical Hardware Organization}

Figure~\ref{fig:arch_overall} presents the hierarchical architecture of Charge-CIM, similar to RAELLA and Cambricon-CIM, consisting of the chip, tile, In-situ Multiply-Accumulator (IMA), Macro array, and MCC levels.
Table~\ref{tab:cc_parameters} summarizes the specific hardware configurations.

\textbf{Chip.} As shown in Figure~\ref{fig:arch_overall}(a), the chip integrates a 4$\times$4  tile array connected by a 2D Mesh NoC, together with global I/O buffers, control logic, and four HyperLink interfaces. These components coordinate data movement, execution timing, inter-tile synchronization, and chip-to-chip communication.

\textbf{Tile.} Each tile contains four IMAs, a 128-KB eDRAM buffer, and a global Special Function Unit (SFU) connected through a high-bandwidth bus (see Figure~\ref{fig:arch_overall}(b)). For execution, DNN layers are partitioned across IMAs, with intermediate partial sums cached in the eDRAM. The Global SFU is optimized for complex operators (e.g., activation, pooling, and batch normalization) and supports multiple concurrent transformations per cycle, effectively eliminating non-linear processing bottlenecks in the pipeline.

\textbf{In-Situ Multiply-Accumulator (IMA).} As shown in Figure~\ref{fig:arch_overall}(c), the IMA integrates a cluster of Macro arrays, a Local SFU, and an I/O buffer. Macro arrays execute parallel MAC operations, with results digitized by differential ADCs. To support intra-IMA layer fusion, a lightweight, Taylor-expansion-based Local SFU is tailored for each output channel. This design enables synchronized, high-throughput processing by matching digital SFU with the analog MAC latency.

\textbf{Macro and Array.} Each Macro comprises two arrays organized as a differential pair  for analog computation and readout (see Figure~\ref{fig:arch_overall}(d)). Within each array, switched-capacitors are reused across data conversion and computation (see Section III).  Consequently, each compute bar (CB) requires only a sense amplifier (SA) and SAR control logic to achieve high-resolution quantization, reducing the converter circuit area.

\textbf{MCC \& Operation.} The MCC achieves a compact footprint by configuring multiple SRAM cells for a single unit capacitor. A NAND-type switch connects the SRAM and the capacitor to perform 1-bit multiplication (see Figure~\ref{fig:arch_overall}(f)). The unit capacitors are implemented in the BEOL layers, allowing vertical stacking above the CMOS logic for higher integration density.  The left plane of Figure~\ref{fig:operation_wave}  shows the six computation   phases within a 20 ns budget ,  including  input charging, DAC conversion, 1-bit multiplication, parallel accumulation, in-situ shift-add (SA), and ADC quantization. Charge is supplied only during input charging and released during ADC quantization.  The accompanying table reports the phase timing and worst-case residual settling error, which is below 0.46 LSB.


\begin{figure}
    \centering
    \includegraphics[width=0.9\linewidth]{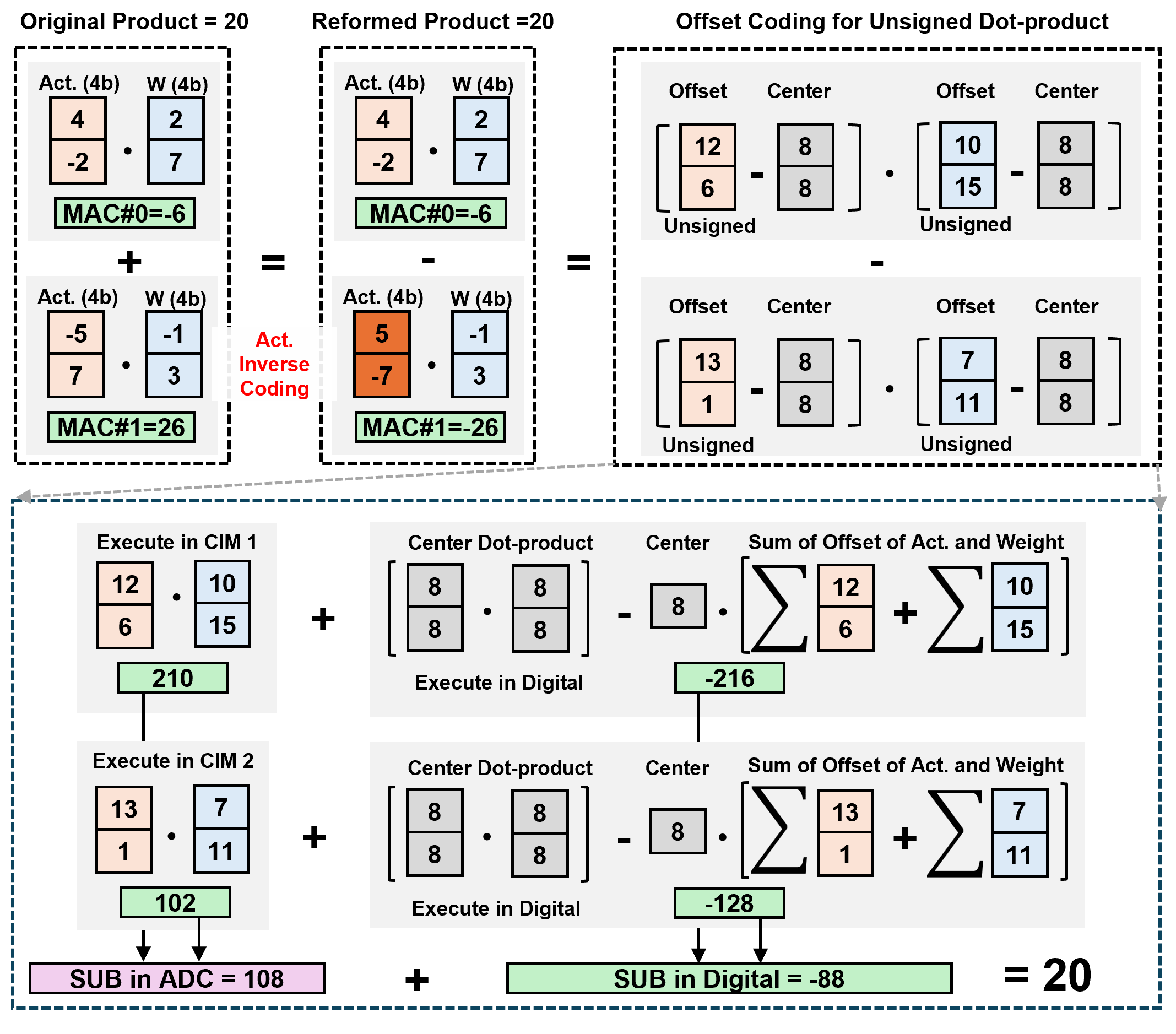}
    \caption{Example of mapping signed dot-products to unsigned CIM operations using Offset and Inverse Coding.}
    \label{fig:toy_example}
\end{figure}

\subsection{In-ADC Psum Add and Inverse Coding}
\label{subsec:diff_adc_coding}

Standard differential SAR-ADCs are widely adopted in data conversion due to their superior common-mode rejection and inherent noise immunity. As illustrated in Figure~\ref{fig:mono_adc}, the interface captures the voltage difference between the positive ($V_P$) and negative ($V_N$) terminals, naturally producing a digital output proportional to the difference: $D_{out} \propto V_P - V_N$. While this physical characteristic inherently performs hardware subtraction, neural network mapping often requires the accumulation of partial sums ($Psum$) distributed across adjacent arrays.

To bridge this gap, we propose an Inverse Coding scheme to realize in-place addition within the subtractive differential interface. Specifically, the $V_N$-side array is driven by pre-inverted inputs ($I_{inv} = -I_N$), generating a negative partial sum $V_N = -I_N \cdot W_N$. In contrast, the $V_P$-side array maintains standard inputs, yielding $V_P = I_P \cdot W_P$. Consequently, the intrinsic subtraction of the differential ADC effectively reconstructs the mathematical sum of the two partial results:
\begin{equation}
    D_{out} \propto V_P - V_N = (I_P \cdot W_P) - (-I_N \cdot W_N) = \sum P_{\mathrm{sum}}
\end{equation}
This mechanism transforms the sensing boundary into an in-situ adder, reducing the number of ADC conversions by half while enhancing the readout resolution to a pseudo-9-bit precision. By leveraging the full differential swing, this strategy significantly improves inference fidelity for large-scale DNNs while reducing both power and area overhead compared to traditional digital accumulation.

\begin{figure}
    \centering
    \includegraphics[width=0.9\linewidth]{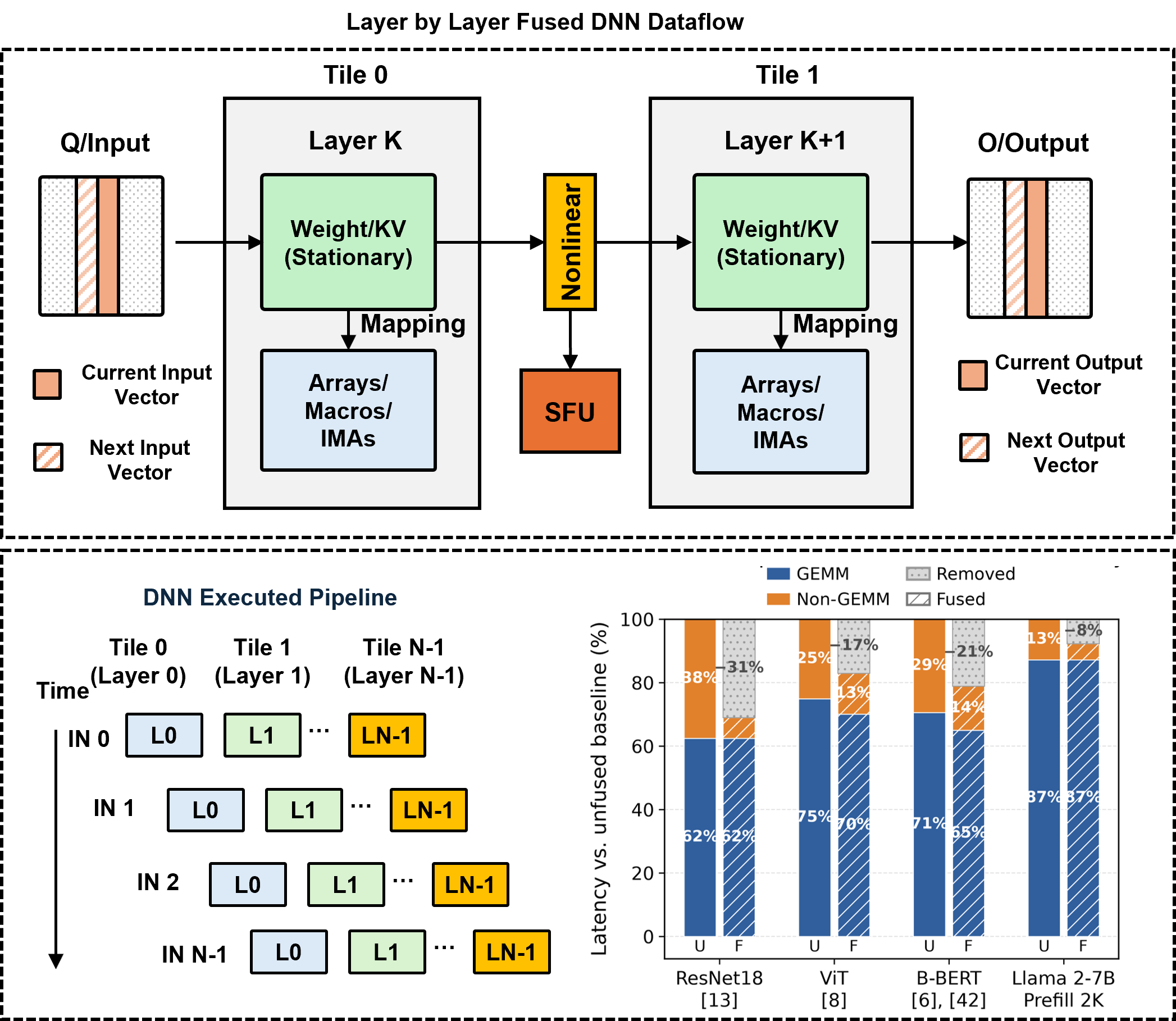}
    \caption{Layer-wise producer--consumer dataflow with stationary weights/KV and overlapped non-GEMM execution.}
    \label{fig:dnn_data_flow}
\end{figure}

\subsection{Joint Offset and Inverse Coding}

Building on the signed arithmetic framework in Section~\ref{subsec:signed_arithmetic}, the joint offset- and inverse-coding workflow is illustrated in Figure~\ref{fig:toy_example}:

\sstab (1) \underline{\textit{Inverse Encoding.}} The dot product is decomposed into two sub-dot products, A and B. The inputs of B are inverse encoded, converting the subtraction operation into a physical differential computation.

\sstab (2) \underline{\textit{Offset Encoding.}} Signed inputs and weights are represented as unsigned offsets relative to a fixed center bias (8 for 4-bit).

\sstab (3) \underline{\textit{Differential Merging.}} The unsigned results of A and B are directly combined in the analog domain by exploiting the subtraction property of differential ADCs.

\sstab (4) \underline{\textit{Constant Compensation.}} The digital backend only handles the linear offset terms. The constant center product term is canceled during differential subtraction, requiring no additional computation.







\subsection{Benefits from Cambricon-CIM}


Charge-CIM natively supports hardware-aware encoding schemes, including the coding-base reformation adopted in Cambricon-CIM, through flexible spatio-temporal mapping of input bit-streams. As shown in Table~\ref{tab:reconfigurable_bitwidth}, non-binary weighting ratios can be realized by reordering the input bits without modifying the capacitor array. For example, an 8-bit input ${B_7, B_6, \dots, B_0}$ can be remapped as ${B_7, B_3, B_2, B_6, B_1, B_0, B_5, B_4}$ to approximate the $3:4$ weighting ratio between $MSB_4$ and $LSB_4$ used in Cambricon-CIM. The resulting effective weights are $(2^2+2^3+2^5+2^6):(2^0+2^1+2^4+2^7)=108:147\approx3:4$. This reconfigurability enables Charge-CIM to support non-uniform encoding and different input precisions.

\subsection{DNN Dataflow}

Figure~\ref{fig:dnn_data_flow} shows a layer-wise producer--consumer pipeline inspired by ISAAC~\cite{shafiee2016isaac} and RAELLA~\cite{andrulis2023raella}. Successive layers are mapped to adjacent compute units, keeping weights/KV stationary while activations stream over the Mesh NoC; pipelining Layer~$K \rightarrow$ SFU $\rightarrow$ Layer~$K{+}1$ overlaps non-GEMM operations with analog MAC.


\section{Evaluation}
\label{sec:evaluation}

\subsection{Experimental Setup}
\begin{table}[t]
\centering
\caption{Hardware specifications and parameters of the Charge-CIM architecture.}
\label{tab:cc_parameters}
\footnotesize
\renewcommand{\arraystretch}{1.2}
\setlength{\tabcolsep}{3pt}

\begin{tabularx}{\columnwidth}{@{} ll c CC @{}}
\toprule
\textbf{Level} & \textbf{Component} & \textbf{Spec.} & \textbf{\makecell[c]{Power \\ (mW)}} & \textbf{\makecell[c]{Area \\ ($mm^2$)}} \\
\midrule

\multirow{2}{*}{\textbf{MCC}}
    & Unit Cap.      & 2.66 fF   & 2.15fJ/act  & 1.74 $\mu m^2$ \\
    & SRAM Cells      & 8        & 0.025 $\mu W$     & 7.1 $\mu m^2$ \\
\midrule

\multirow{3}{*}{\textbf{Macro}}
    & MCC Array      & 256$\times$128$\times$8b & 28.15 & 1.43 \\
    & Input Drivers  & 256$\times$8b & 0.12  & 0.033 \\
    & Embedded ADC   & 128$\times$8b     & 18.71 & 0.069 \\
\midrule

\multirow{3}{*}{\textbf{IMA}}
    & Macro Group    & 2 Macros & 93.94 & 3.06\\
    & Local SFU      & 128$\times$ PE-based & 13.40 & 0.07 \\
    & I/O Buffer     & 4 KB SRAM          & 2.90  & 0.025 \\
\midrule

\multirow{3}{*}{\textbf{Tile}}
    & IMA Group     & 4 IMAs             & 440.96 & 12.64\\
    & Global SFU     & 32                & 32.31  & 0.08 \\
    & eDRAM          & 128 KB             & 28.98 & 0.320 \\
    & Tile Bus     & 256b             & 3.2  & 5.62 \\
\midrule

\multirow{3}{*}{\textbf{Chip}}
    & Tile Group     & 16 Tiles & 8087 &  298.50 \\
    & HyperLink     & 1/1.6 GHz       & 742.5  & 1.07 \\
    & Link BW        & 6.4 GB/s        & --     & -- \\
    
Total & \textbf{--} & \textbf{--}
& \makecell[c]{ 8.8 W (Peak) \\ 4.5 W (Avg.)\textsuperscript{a}}
& 299.57 \\

\bottomrule
\multicolumn{5}{l}{\scriptsize \textsuperscript{a} Average power estimated by architecture-level simulation at 1 GHz.}
\end{tabularx}
\end{table}



\begin{table}[t]
\centering
\caption{Analog MAC mismatch analyzed by Monte Carlo simulation under different capacitor sizes.\textsuperscript{a}}
\label{tab:cap_mc_error}

\footnotesize
\setlength{\tabcolsep}{10pt}
\renewcommand{\arraystretch}{1.2}
\begin{tabular}{@{}cccccc@{}}

\hline
\multirow{2}{*}{\makecell[c]{Cap.\\(fF)}} &
\multirow{2}{*}{\makecell[c]{Area\\($\mu\mathrm{m}^{2}$)}} &
\multicolumn{2}{c}{Mismatch (mV)} &
\multicolumn{2}{c}{Mismatch (LSB)} \\
\cmidrule(lr){3-4}\cmidrule(lr){5-6}
& &
$1\sigma$ &
$3\sigma$ &
$1\sigma$ &
$3\sigma$ \\
\hline
1.460 & 1.35&3.372 & 10.115 & 0.480 & 1.438 \\
2.06&  1.89&3.087 & 9.260  & 0.439 & 1.317 \\
\textbf{2.66}&  2.43&\textbf{2.775} & \textbf{8.326} & \textbf{0.395} & \textbf{1.184} \\
3.26&  2.97&2.485 & 7.456  & 0.354 & 1.060 \\
3.86&  3.51&2.234 & 6.702  & 0.318 & 0.953 \\
\hline

\multicolumn{6}{@{}l@{}}{%
    \scriptsize\textsuperscript{a} 1.8 V P-P dynamic range in differential INT-8 CIM.
}
\end{tabular}
\end{table}

\begin{figure}
    \centering
    \includegraphics[width=0.9\linewidth]{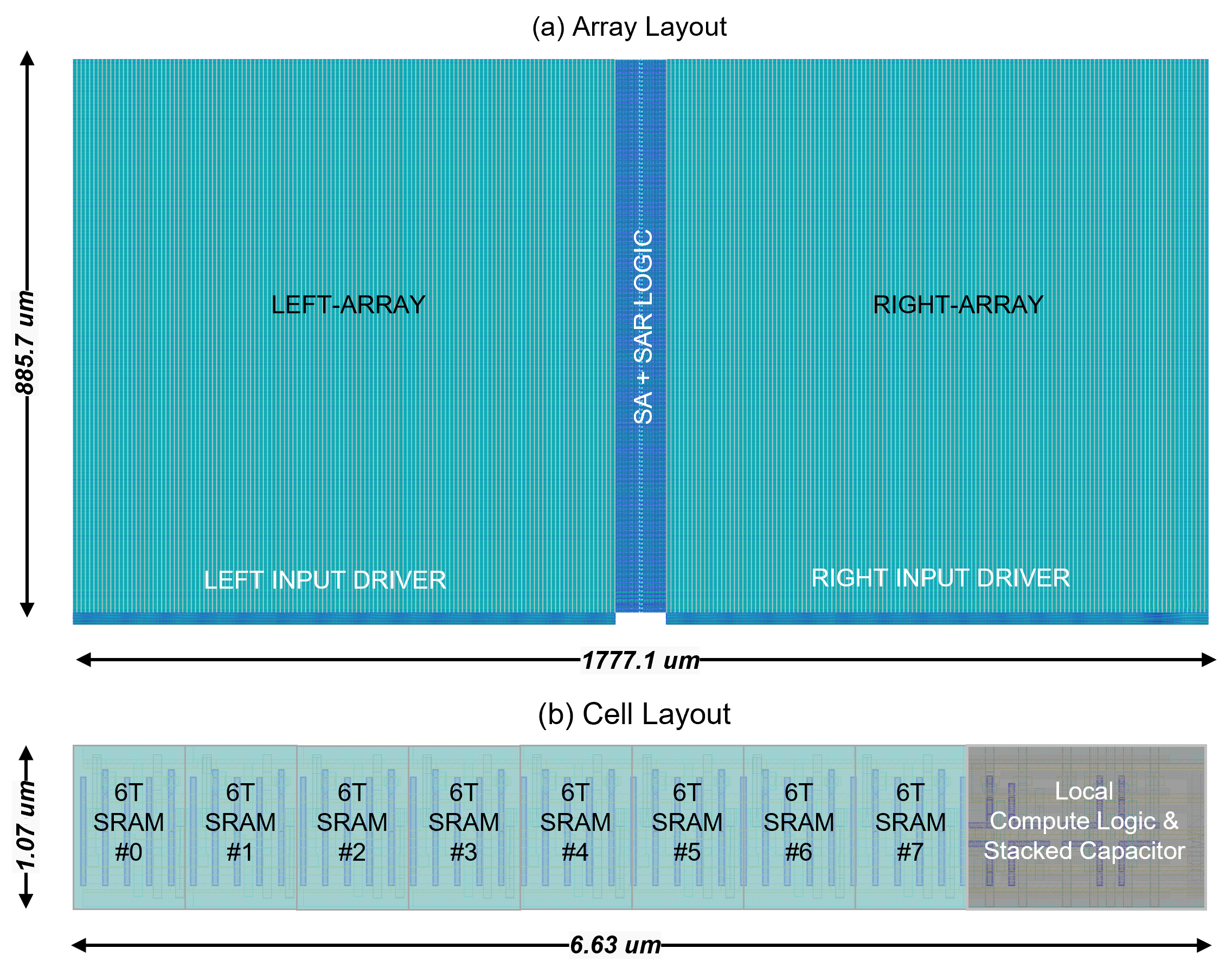}
    \caption{Post-layout implementation used for circuit simulation: (a) full macro layout and (b) cell-level layout.}
    \label{fig:array_cell_layout}
\end{figure}

\textbf{Hardware Configurations.} For accurate evaluation, we adopt a cross-layer simulation methodology following Cambricon-CIM. All key analog circuits, including the MCC array and embedded ADC, are simulated in SPICE under 28-nm CMOS process. Their power, latency, area, and accuracy are extracted from post-layout simulations. For non-ideality modeling, the Charge-CIM datapath is divided into three stages: embedded DAC conversion, analog VMM, and embedded ADC quantization. The embedded DAC and analog MAC errors are independently extracted from circuit simulations and modeled as zero-mean Gaussian variations $\sigma_{\mathrm{DAC}}^{2}$ and $\sigma_{\mathrm{MAC}}^{2}$. Together with the analytically estimated $kT/C$ thermal noise~\cite{gonugondla2021fundamental}, the total analog voltage noise variance is:
\begin{equation}
\sigma_{\mathrm{noise}}^{2}
=
\sigma_{\mathrm{DAC}}^{2}
+
\sigma_{\mathrm{MAC}}^{2}
+
\frac{kT}{2mC_u}
\left[
1+\frac{1}{n}\left(1+\frac{1}{m}\right)
\right]
\end{equation}
where $k$ is the Boltzmann constant, $T$ is the absolute temperature, $m$ and $n$ are the numbers of columns and rows, respectively, and $C_u$ is the unit capacitance. As shown in Table~\ref{tab:cap_mc_error}, $C_u=2.66$ fF balances capacitor mismatch and area overhead, limiting the $3\sigma$ variation to approximately 1.18 LSB.
The ADC quantization error is modeled by rounding the partial sums according to the 8-bit ENOB extracted from ADC simulations. To improve density, unit capacitors are stacked above the switch-control logic, as shown in Figure~\ref{fig:array_cell_layout}(b). The $256 \times 128$ macro is split into two $128 \times 128$ subarrays operating differentially for improved noise immunity. The digital modules, including the SFUs and controller, are synthesized at 1 GHz using Synopsys Design Compiler and analyzed with PrimeTime. Nonlinear functions are evaluated through iterative Taylor-series expansion to match the 20 ns analog VMM latency. The area and power of SRAM, eDRAM, and interconnects are obtained using CACTI~7.0~\cite{balasubramonian2017cacti}. All component-level parameters are integrated into the Timeloop/Accelergy~\cite{parashar2019timeloop,wu2019accelergy,andrulis2024cimloop} framework to evaluate performance and power across nine DNN benchmarks. Table~\ref{tab:cc_parameters} summarizes the Charge-CIM hardware parameters.

\textbf{Hardware Baseline.} We compare Charge-CIM with RAELLA~\cite{andrulis2023raella} and Cambricon-CIM~\cite{guo2026cambricon}, which employ Center + Offset encoding and coding-base reformation, respectively, to support 7-bit or 6-bit ADC operation. For a fair comparison, all designs are normalized to a 300 mm$^2$ area budget, with RAELLA configured using SRAM. Cambricon-CIM and RAELLA are scaled to 20 and 18 tiles, respectively, each containing 16 arrays of size $72\times80\times4$\ b.

\textbf{DNN Models and Datasets.} We evaluate nine representative models: VGG16 \cite{simonyan2014very}, ResNet18 \cite{he2016deep}, MobileNetV3 (MbNV3) \cite{Howard2019SearchingFM}, DenseNet201 (DN201) \cite{Huang2016DenselyCC}, and Vision Transformer (ViT) \cite{Dosovitskiy2020AnII} on ImageNet classification; MobileBERT (M-BERT) \cite{Sun2020MobileBERTAC} and Base-BERT (B-BERT) \cite{Devlin2019BERTPO, Wu2020IntegerQF} on Stanford Question Answering Dataset (SQuAD); and GPT-2 Medium \cite{Radford2019LanguageMA} and Llama 2-7B \cite{Touvron2023Llama2O} on WikiText-2, focusing on LLM prefill. All weights and activations are quantized to INT8 using PyTorch/TorchVision without retraining.

\subsection{Accuracy}

\begin{figure}
    \centering
    \includegraphics[width=1\linewidth]{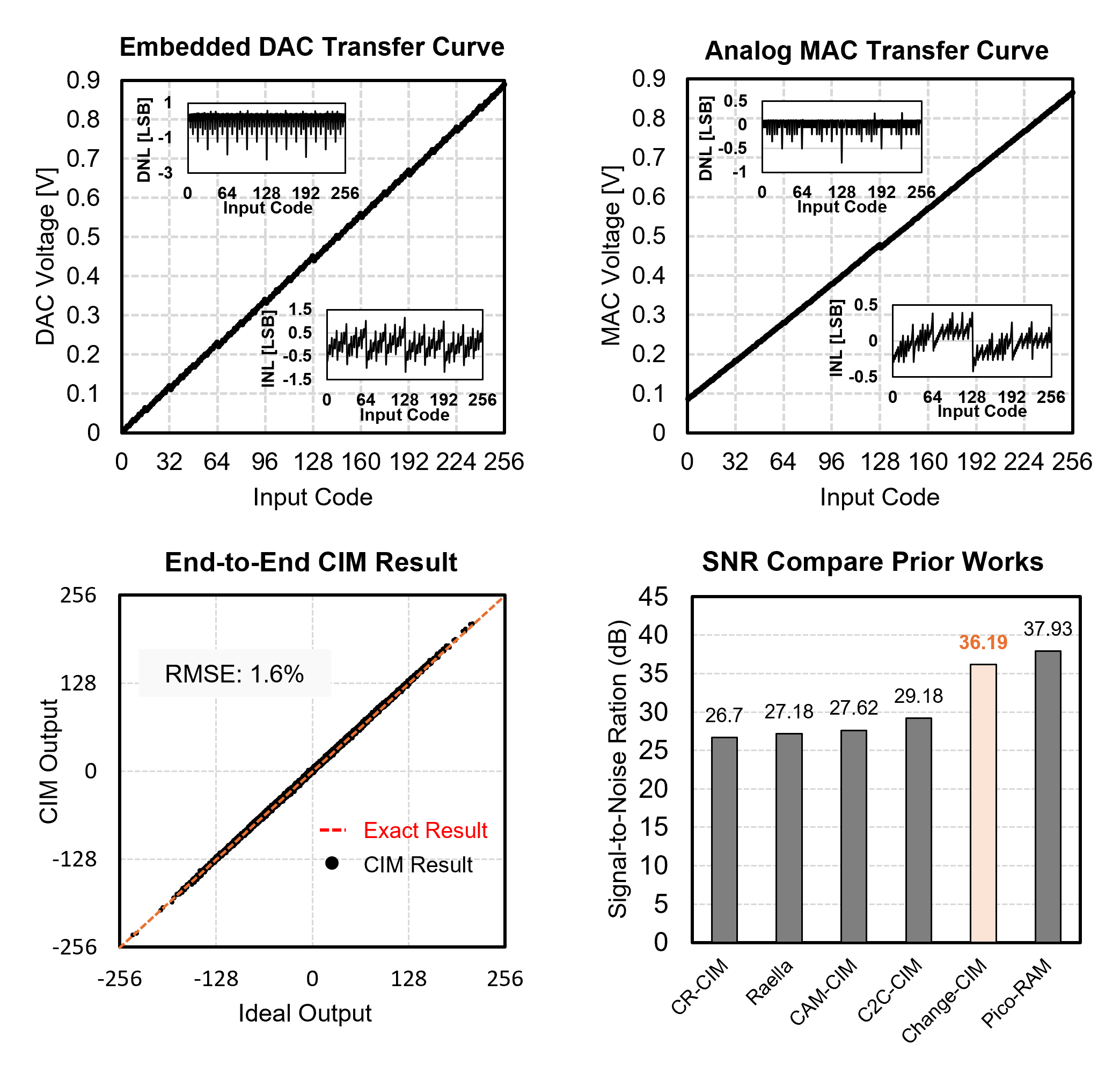}
    \caption{Evaluation of Charge-CIM accuracy and robustness: (a) embedded-DAC transfer curve and INL/DNL, (b) analog-MAC transfer curve and INL/DNL, (c) end-to-end CIM error, and (d) SNR comparison with prior CIM designs.}
    \label{fig:error_analysis}
\end{figure}

\textbf{Analog VMM Error.} Following the non-ideality modeling flow described above, we first extract the transfer characteristic of the embedded DAC. Figure~\ref{fig:error_analysis}(a) shows a differential non-linearity (DNL) of $+0.52/-2.24$~LSB and an integral non-linearity (INL) of $+1.12/-1.14$~LSB. The analog MAC transfer characteristic is then evaluated in Figure~\ref{fig:error_analysis}(b), showing a DNL of $+0.15/-1.64$~LSB and an INL of $+0.86/-0.79$~LSB. By incorporating the extracted DAC and MAC errors, ADC quantization, and thermal noise, we compare the ideal and simulated CIM outputs in Figure~\ref{fig:error_analysis}(c), yielding an end-to-end VMM root-mean-square error (RMSE) of $1.6\%$. Finally, Figure~\ref{fig:error_analysis}(d) shows that Charge-CIM achieves a signal-to-noise ratio (SNR) of $36.19$~dB, approximately $9$~dB higher than RAELLA and Cambricon-CIM.

\textbf{DNN Inference Accuracy.} As summarized in Table~\ref{tab:quant_presicion}, the modeled hardware non-idealities change vision-model accuracy by at most 1.9 percentage points relative to the reported FP32 results, while the BERT variants lose at most 1.7 points in F1 and 1.4 points in exact match. Llama 2-7B and GPT-2 exhibit PPL increases of 0.26 and 2.9, respectively.


\begin{table}[t]
\centering
\caption{Accuracy/performance degradation under progressive hardware non-idealities.}
\label{tab:quant_presicion}
\footnotesize
\setlength{\tabcolsep}{3.5pt}
\renewcommand{\arraystretch}{1.2}

\begin{tabular}{@{} l ccccc @{}}
\toprule
\textbf{Model} & \textbf{FP32} & \textbf{INT8} & \textbf{Analog\textsuperscript{a}} & \textbf{ADC\textsuperscript{b}} & \textbf{Final Loss} \\
\midrule
\multicolumn{6}{@{}l}{\textbf{ImageNet (Top-1 Accuracy \% $\uparrow$)}} \\
\cmidrule{1-6}
VGG16      & 86.5 & 86.5 & 86.6 & 86.4 & -0.1 \\
Res18      & 77.8 & 80.3 & 80.6 & 79.7 & +1.9 \\
MbV3       & 86.8 & 86.5 & 86.6 & 86.5 & -0.3 \\
DN201      & 88.5 & 88.1 & 88.0 & 87.7 & -0.8 \\
ViT        & 85.0 & 85.0 & 84.9 & 85.1 & +0.1 \\
\midrule
\multicolumn{6}{@{}l}{\textbf{SQuAD (F1 / EM)}} \\
\cmidrule{1-6}
M-BERT     & 90.0/82.9 & 89.9/82.7 & 89.3/82.1 & 88.5/81.5 & -1.5/-1.4 \\
B-BERT   & 88.5/81.2 & 88.3/81.0 & 87.6/80.7 & 86.8/79.8 & -1.7/-1.4 \\
\midrule
\multicolumn{6}{@{}l}{\textbf{WikiText-2 (PPL $\downarrow$)}} \\
\cmidrule{1-6}
GPT-2      & 22.8 & 23.0 & 23.6 & 25.7 & +2.9 \\
Llama 2-7B & 5.47 & 5.51 & 5.52 & 5.73 & +0.26 \\
\bottomrule
\multicolumn{6}{@{}p{\columnwidth}@{}}{\scriptsize \textsuperscript{a} Modeled with RMSE = 1.6\% charge redistribution variation in Charge-CIM arrays; \textsuperscript{b} Evaluated with 8-bit ENOB SAR-ADC.}
\end{tabular}
\end{table}

\begin{figure*}
    \centering
    \includegraphics[width=1\linewidth]{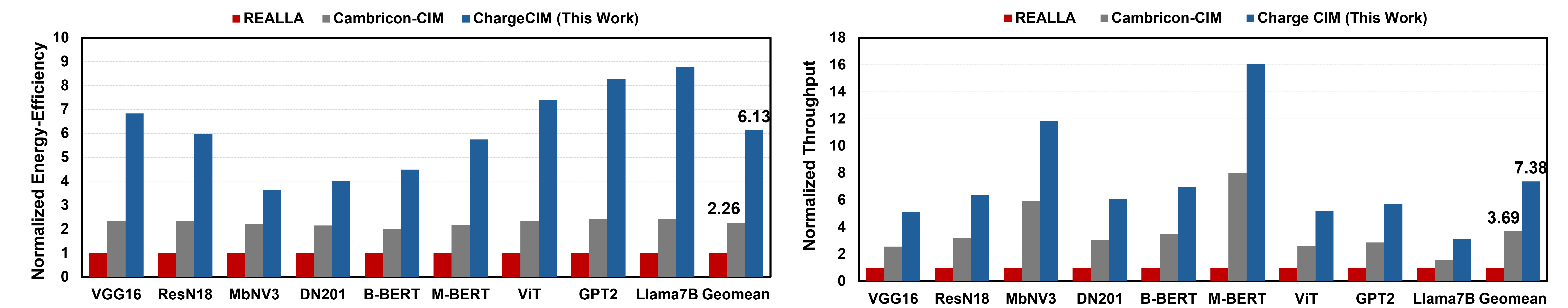}
    \caption{Energy efficiency and throughput normalized to RAELLA across nine DNN models. Cambricon-CIM achieves geometric means of 2.26$\times$ and 3.69$\times$, respectively; Charge-CIM achieves 6.13$\times$ and 7.38$\times$, corresponding to 2.71$\times$ higher energy efficiency and 2.00$\times$ higher throughput than Cambricon-CIM.}
    \label{fig:normalized_ee_throughput}
\end{figure*}

\begin{figure}
    \hspace{0.02\linewidth}
    \includegraphics[width=0.98\linewidth]{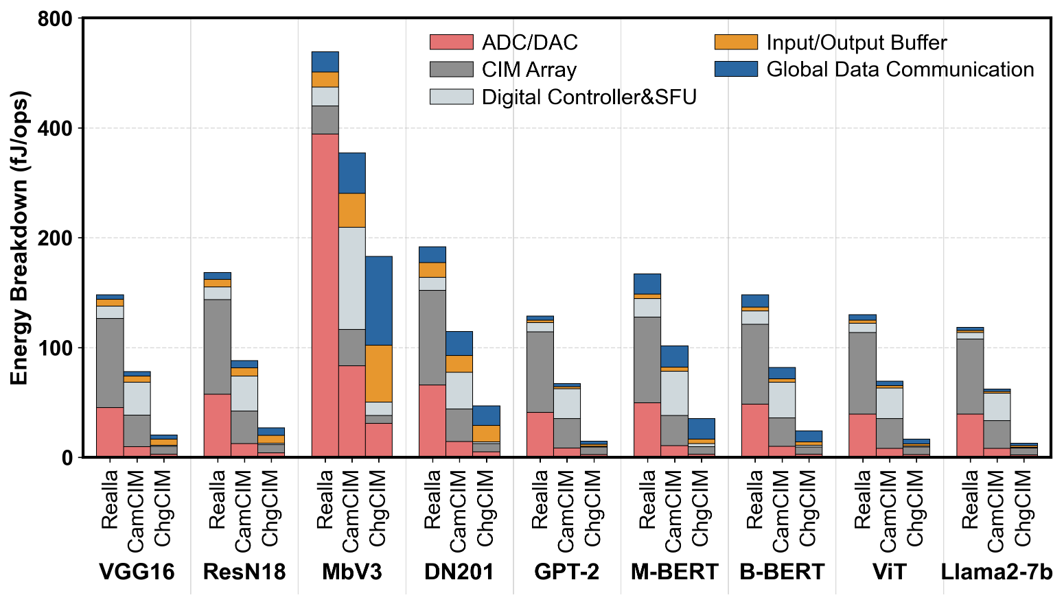}
    \caption{Energy breakdown of RAELLA, Cambricon-CIM (CamCIM), and Charge-CIM (ChgCIM) across the evaluated neural-network models.}
    \label{fig:energy_breakdown}
\end{figure}

\subsection{Energy Efficiency}


We evaluate the energy efficiency of Charge-CIM across all DNN benchmarks, comparing it against SOTA SRAM-based charge-domain CIM architectures, including Cambricon-CIM and RAELLA (as the normalized baseline). As shown in Figure~\ref{fig:normalized_ee_throughput} left, Charge-CIM achieves a $3.6\times$--$8.8\times$ energy efficiency improvement over RAELLA, with a geometric mean of $6.13\times$. Compared to Cambricon-CIM, our design still maintains a $1.7\times$--$3.6\times$ advantage ($2.7\times$ on average).

The large energy-efficiency gains, especially for GPT-2 (8.27$\times$) and Llama 2-7B (8.77$\times$), mainly come from avoiding intermediate conversions across the input and weight bit slices used by RAELLA and Cambricon-CIM. Those designs require multiple cycles, ADC conversions, and digital reconstruction for each multi-bit output. Charge-CIM instead combines the bit-weighted partial results through in-situ charge redistribution before one output conversion per compute-bar read. Figure~\ref{fig:energy_breakdown} compares the normalized energy breakdown in fJ/op. Bit-parallel analog accumulation and the embedded ADC reduce conversion energy, while avoiding the digital encoding and reconstruction overhead of Cambricon-CIM, which contributes up to 26.2\% of total energy. Charge-CIM also avoids the more than 2.5$\times$ array-energy overhead introduced by weight-centering encoding.


\begin{figure}
    \centering
    \includegraphics[width=0.8\linewidth]{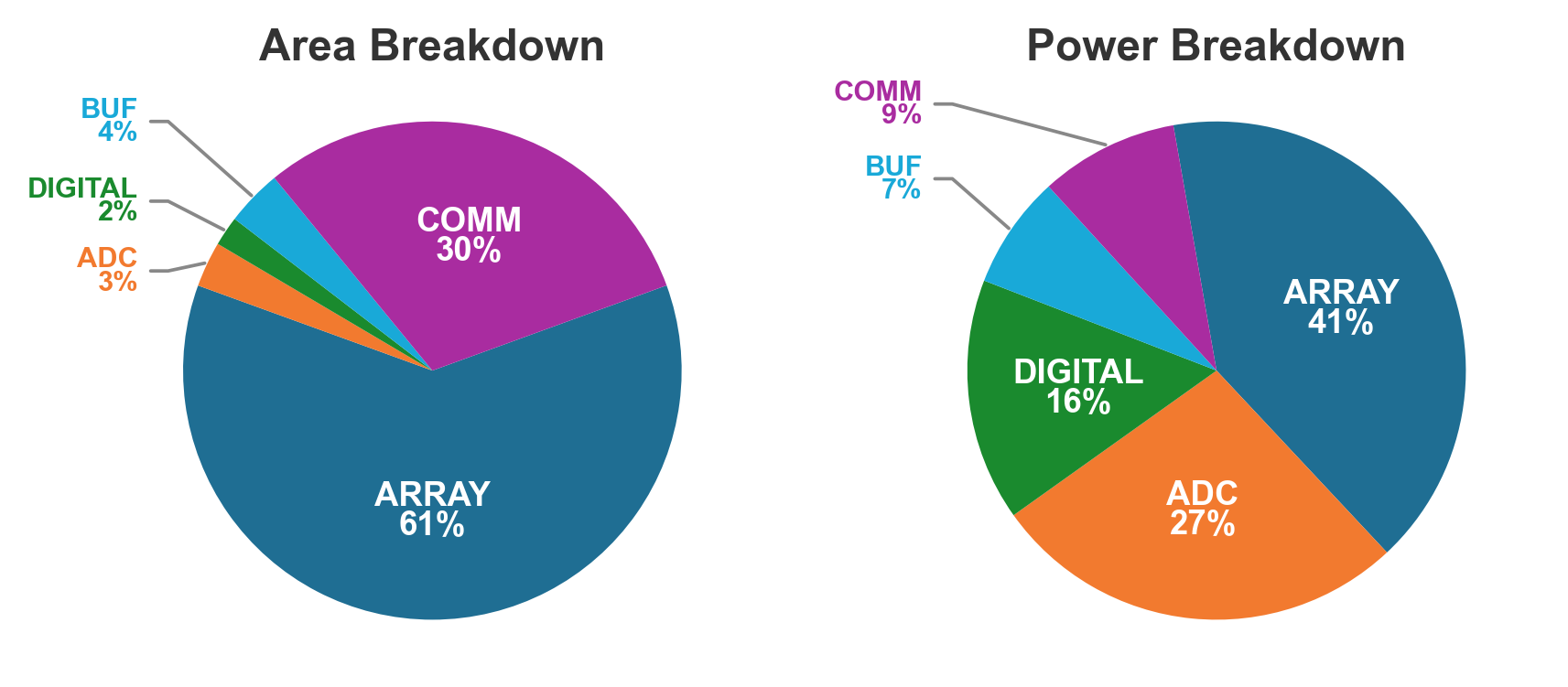}
    \caption{Area and power breakdown.}
    \label{fig:pa_breakdown}
\end{figure}

\subsection{Throughput}


Figure~\ref{fig:normalized_ee_throughput} right compares the performance of Charge-CIM against RAELLA and Cambricon-CIM. Under equivalent area constraints, Charge-CIM achieves a throughput speedup ranging from $3.0\times$ to $16.0\times$ (averaging $7.3\times$) across various DNN models, outperforming Cambricon-CIM by $2.0\times$ on average. Finally, our architecture delivers a high area efficiency of $1.4 \text{ TOPS/mm}^2$ at the system level.


These gains stem from bit-parallel analog 8-bit VMM, which avoids the input- and weight-bit serialization and intermediate conversions required by the baseline designs. Charge-CIM further employs compact per-channel embedded ADCs to maximize readout parallelism, whereas RAELLA and Cambricon-CIM rely on ADC multiplexing to reduce area overhead. As a result, even with 7-bit ADCs, these baselines require approximately 100 ns per array readout. Moreover, the compact differential array increases the number of compute units within a given area while doubling the quantization range to $V_{PP}=2V_{REF}$. This improves resolution and reduces the additional ADC activations caused by matrix partitioning, enabling higher computational density and throughput.



\subsection{Power and Area Breakdown}

Figure~\ref{fig:pa_breakdown} shows the area and power breakdown of the proposed Charge-CIM at the architectural level. The array dominates the area footprint, contributing 61\% of the total area, followed by communication (COMM) at 30\%. Other components, including ADC, on-chip buffer (BUF), and digital logic, account for only 3\%, 4\%, and 2\%, respectively. Power is dominated by the arrays and ADCs at 41\% and 27\%, followed by the digital subsystem at 23\% and communication at 9\%.

\begin{figure}
    \centering
    \includegraphics[width=0.95\linewidth]{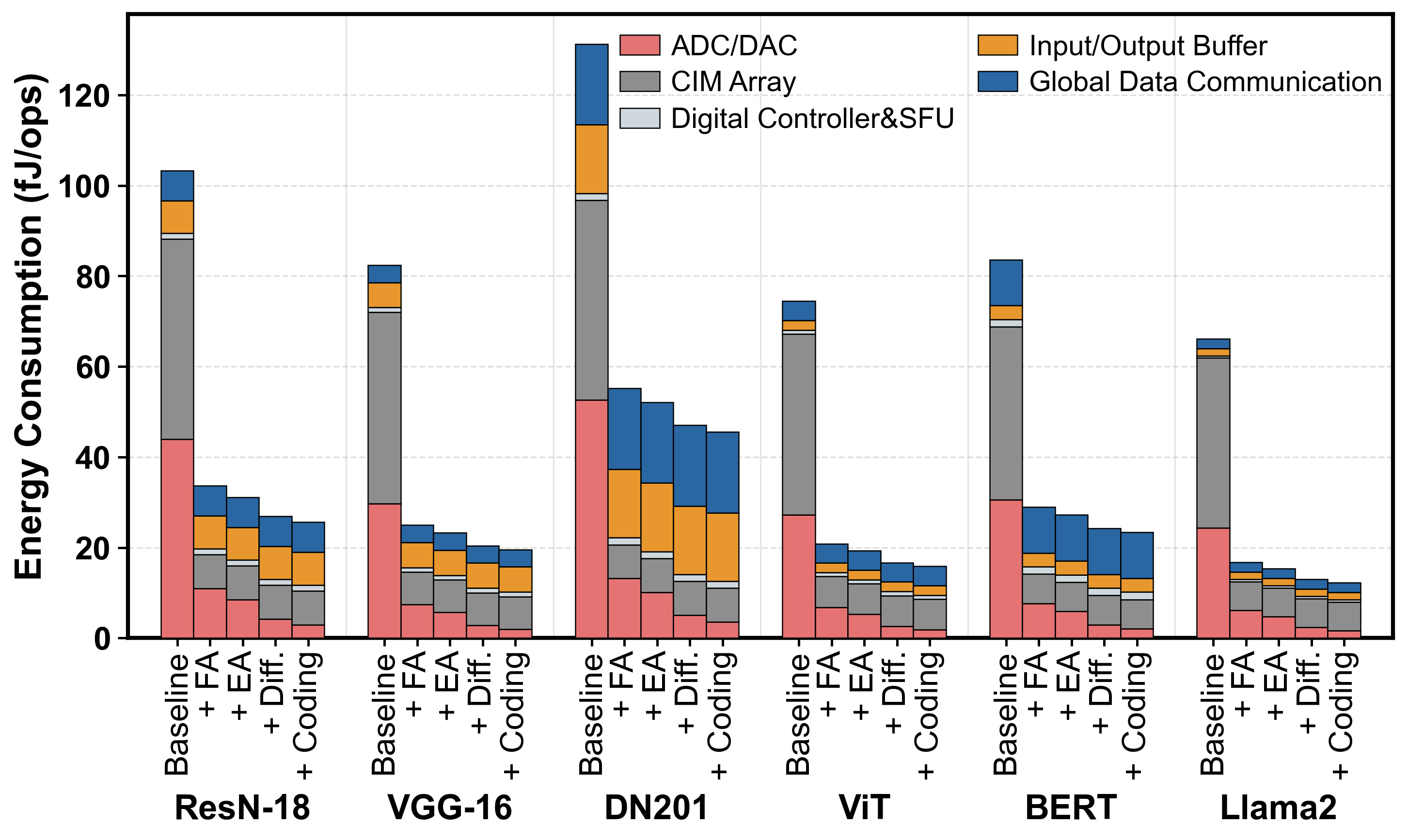}
    \caption{Ablation study of energy consumption across different DNN models, showing progressive reductions from fully analog computing (FA), embedded ADC (EA), differential array (Diff.), and base-reforming coding.}
    \label{fig:ablation_final_labels}
\end{figure}

\subsection{Ablation Study}






As shown in Figure~\ref{fig:ablation_final_labels}, we use the Cambricon-CIM array and bit-width configuration as the baseline and progressively introduce fully analog computing (FA), an embedded ADC (EA), a differential array (Diff.), and base-reforming coding to quantify their individual contributions.

\sstab (1) \underline{\textit{Fully Analog Computing (FA).}} FA replaces conventional bit-sliced computation, reducing array energy by $5.9\times$. By eliminating repeated intermediate conversions, it also lowers ADC energy by $4\times$.

\sstab (2) \underline{\textit{Embedded ADC (EA).}} The customized embedded ADC shortens the quantization path and reduces 25\% ADC energy.

\sstab (3) \underline{\textit{Differential Array (Diff.).}} The differential structure improves noise tolerance and fully utilizes the voltage swing, providing an additional $2\times$ energy reduction.

\sstab (4) \underline{\textit{Base-Reforming Coding.}} Adopting the Cambricon-CIM coding scheme enables lower-resolution ADCs and further reduces the associated overhead by approximately 12\%.

Overall, these optimizations reduce the total conversion overhead by more than 91.7\%, delivering consistent energy savings across all evaluated DNN benchmarks.

\subsection{Robustness Analysis}

\textbf{VMM Shape.} Table~\ref{tab:vmm_performance} demonstrates the scalability of the Charge-CIM macro across different VMM dimensions. For shapes up to 256, it achieves excellent linearity ($R^2 > 0.9999$) with a maximum error below 0.7\%. Larger workloads are spatially partitioned across sub-arrays, with partial sums accumulated digitally. Therefore, analog errors do not accumulate, bounding the overall maximum error to approximately 0.7\%, corresponding to the single-array worst case.

\begin{figure}
    \centering
    \includegraphics[width=1\linewidth]{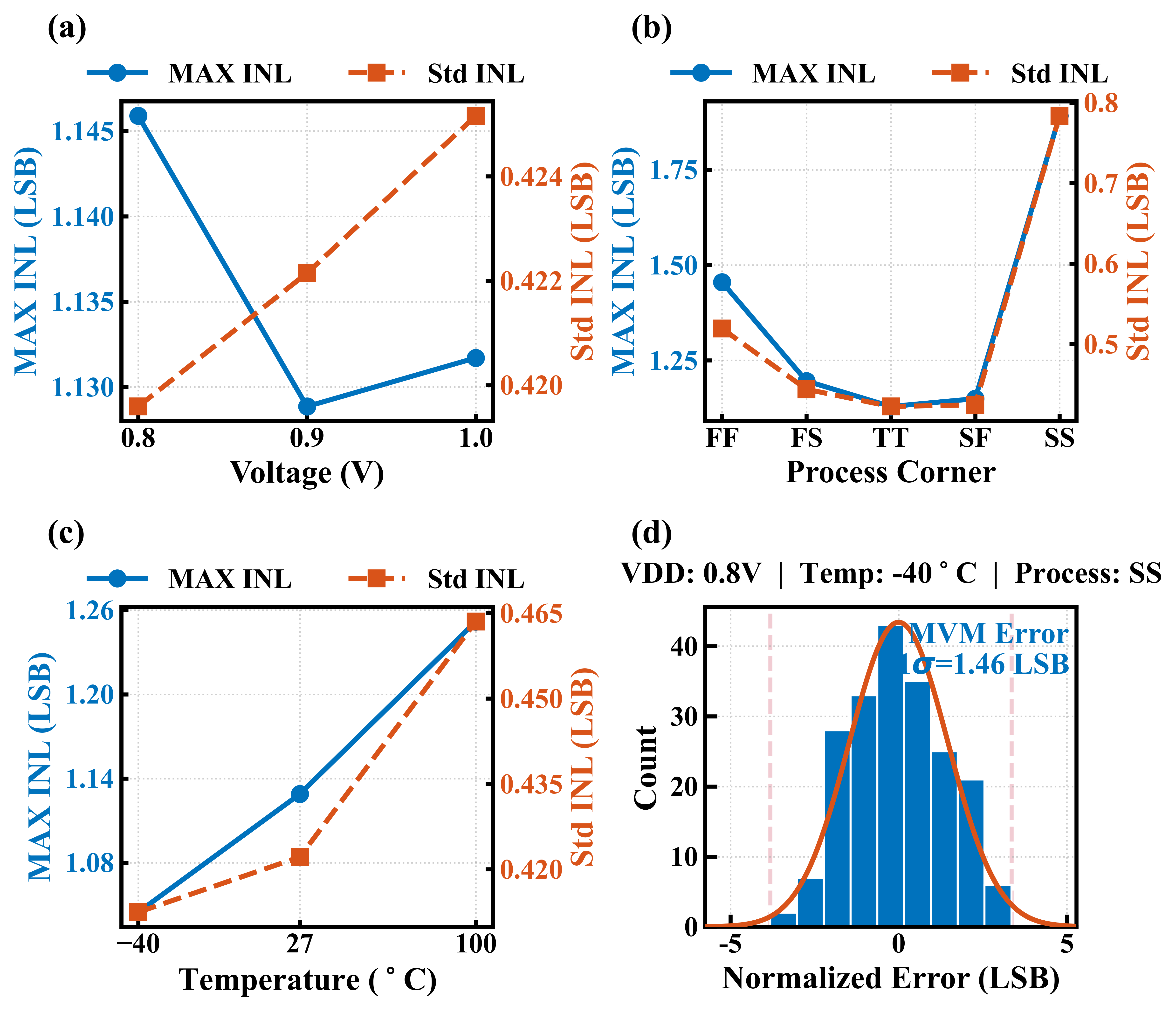}
    \caption{Post-layout robustness evaluation of Charge-CIM: (a) supply-voltage variation, (b) process variation, (c) temperature variation, and (d) Monte-Carlo normalized MVM error distribution under the worst-case corner.}
    \label{fig:pvt}
\end{figure}
\vspace{-1mm}

\begin{table}[htbp]
\centering
\caption{Linearity and error performance under different VMM shapes.}
\label{tab:vmm_performance}
\resizebox{\columnwidth}{!}{%
\begin{tabular}{lcccc}
\toprule
\textbf{VMM Shape} & \textbf{Shape $=64$} & \textbf{Shape $=128$} & \textbf{Shape $=196$} & \textbf{Shape $\geq$ 256} \\
(Array Utilization) & (25\%) & (50\%) & (76.6\%) & (100\%) \\
\midrule
$R^2$ (Goodness of Fit) & 0.9999 & 0.9999 & 0.9999 & 0.9999 \\
Max Linearity Error (\%) & 0.24 &  0.34 &  0.42 & 0.70 \\
\bottomrule
\end{tabular}%
}
\end{table}

\textbf{PVT and Monte Carlo Robustness.} Post-layout simulations are conducted across TT, SS, FF, FS, and SF corners, with temperatures ranging from $-40^\circ\mathrm{C}$ to $100^\circ\mathrm{C}$ and a supply voltage of $0.9~\mathrm{V} \pm 10\%$. Monte Carlo simulations further evaluate local capacitor and transistor mismatches. As shown in Figure~\ref{fig:pvt}, the maximum computational error remains below 2 LSB, while the error variance stays below 1.46 LSB even at the worst process corner (0.8V, $-40^\circ\mathrm{C}$, SS corner), demonstrating the robustness of the proposed architecture.

\begin{figure}
    \centering
    \includegraphics[width=1\linewidth]{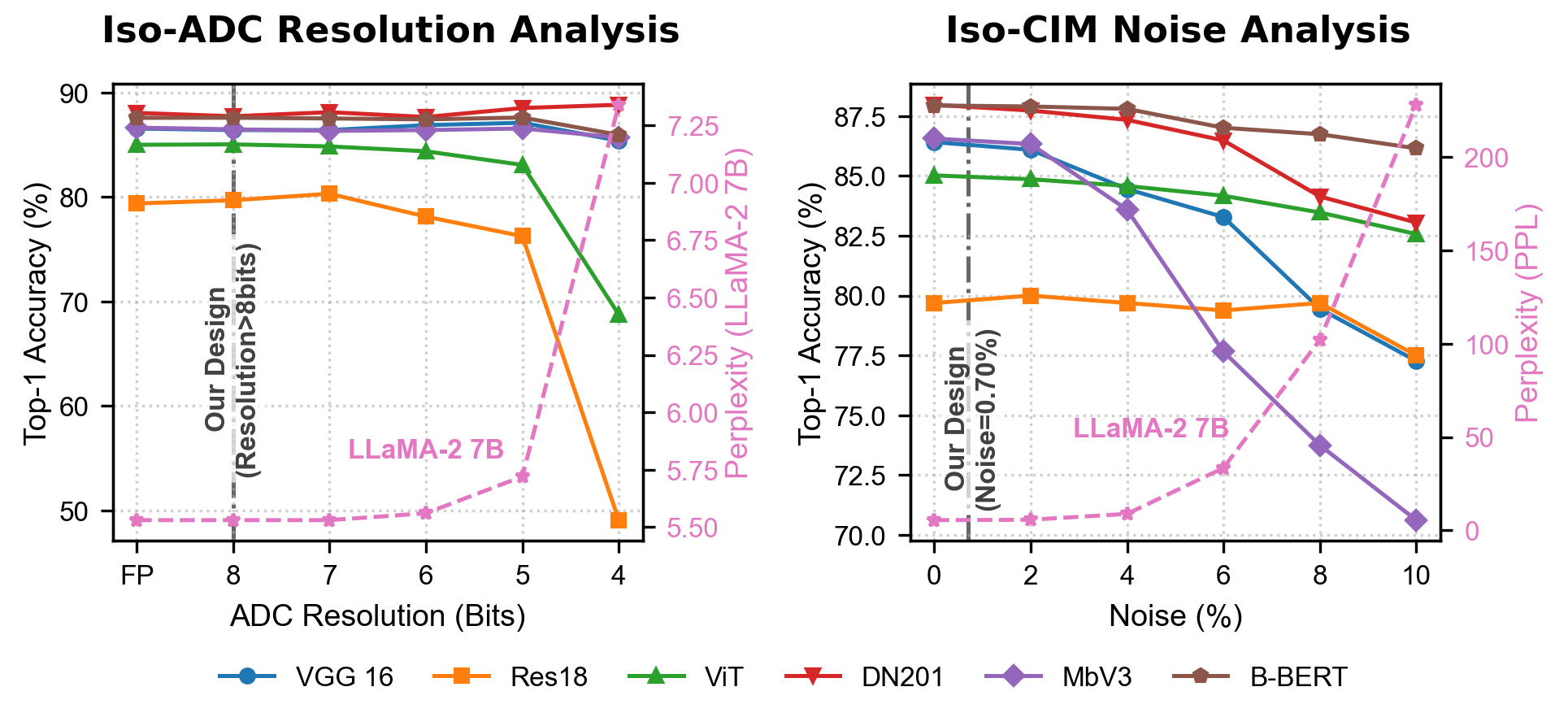}
    \caption{DNN and LLM inference robustness under analog CIM non-idealities: (a) reduced ADC resolution and (b) increased analog CIM noise.}
    \label{fig:dnn_robustness}
\end{figure}

\textbf{DNN Inference Robustness.} Figure~\ref{fig:dnn_robustness} evaluates the impact of ADC resolution and analog noise on inference accuracy. As shown in Figure~\ref{fig:dnn_robustness}(a), most models remain stable down to 5-bit ADC resolution and degrade at 4 bits, providing a substantial precision margin for the adopted 8-bit ADC. Figure~\ref{fig:dnn_robustness}(b) further shows strong robustness under 2\% analog noise. Since the simulated MAC error is only 0.70\%, Charge-CIM operates well within this robust regime and maintains high inference accuracy across different models without retraining.

\section{Related Work}

\label{sec:related}

\begin{figure}
    \centering
    \includegraphics[width=1\linewidth]{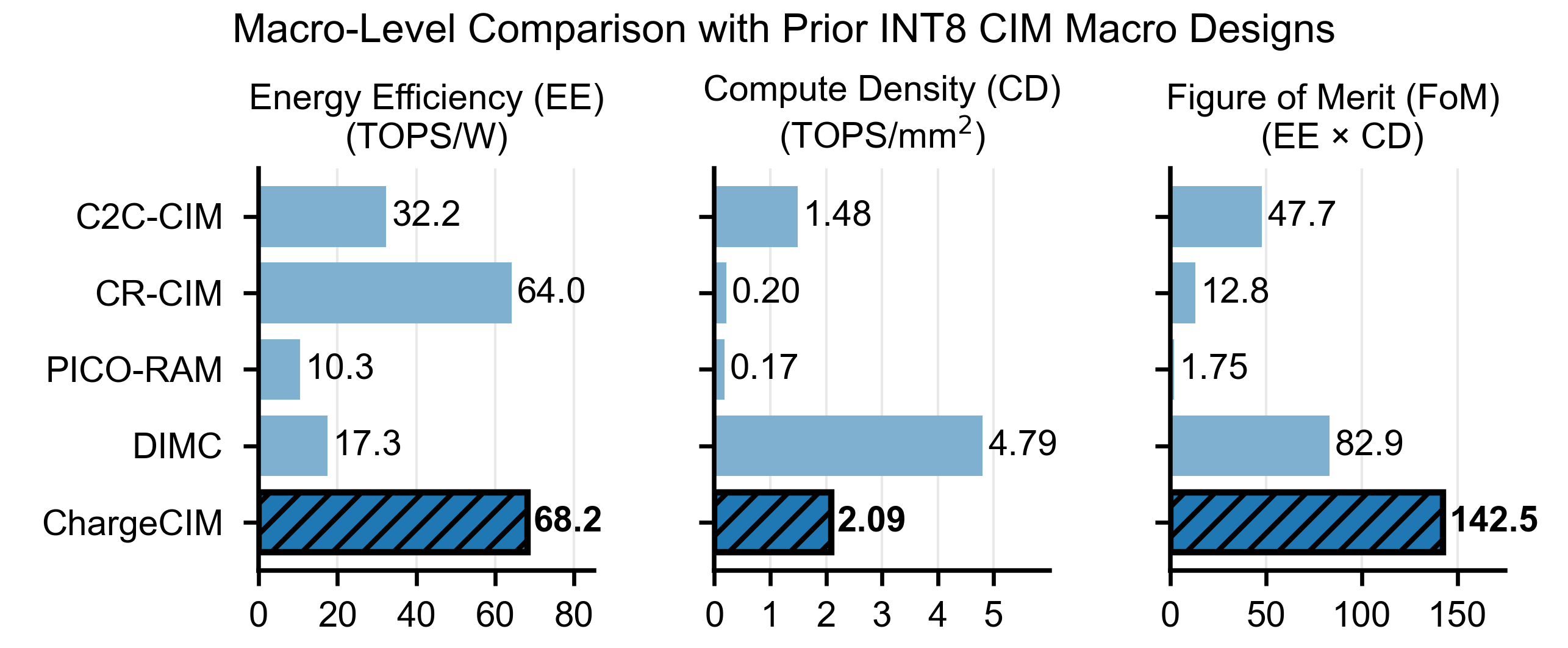}
    \caption{Macro-level comparison of Charge-CIM with state-of-the-art INT8 charge-domain CIM macros (C2C-CIM~\cite{wang2023charge}, CR-CIM~\cite{yoshioka2024818}, and PICO-RAM~\cite{chen2024pico}) and the digital CIM macro DIMC~\cite{wang2022dimc} in terms of energy efficiency (EE), compute density (CD), and their figure-of-merit (FoM).}
    \label{fig:cim_macro_comparsion}
\end{figure}

\textbf{Charge-Domain CIM Technologies.} Charge-domain CIM is attractive for its high linearity, PVT robustness, and energy efficiency. Prior works have explored different circuit optimizations: CAP-RAM~\cite{chen2021cap} achieves 0.9999 computing linearity through bitline-input charge processing; Dong et al.~\cite{dong202015} employ a complementary capacitor ladder for 4-bit computing; PICO~\cite{chen2024pico} improves density with compact cells; and Xie et al.~\cite{Xie2021162EC} integrate eDRAM-based charge computing and adaptive conversion. More recent designs reduce peripheral overhead through C-2C ladders~\cite{wang2023charge} and circuit reuse~\cite{xuan2025yoco,lee202328}, but still depend on costly ADC readout. CR-CIM~\cite{yoshioka2024818} lowers ADC overhead but requires extensive bit-slicing due to its 1-bit capacitive operation. Figure~\ref{fig:cim_macro_comparsion} compares our design with state-of-the-art INT8 charge-domain and digital CIM macros~\cite{wang2022dimc} in energy efficiency and compute density.

\textbf{Co-Design CIM Architecture.} Beyond circuit optimization, architectural co-design is essential for scalability and throughput. Prior works improve programmability and communication through reconfigurable NoCs~\cite{jia202115}, analog buffers~\cite{li2020timely}, coding-based ADC reduction~\cite{guo2024cambricon}, pipelined execution~\cite{song2017pipelayer}, and sparsity support~\cite{qi2025ciminus}.

\section{Conclusion}


Charge-CIM addresses the ADC wall by replacing bit-sliced ACiM execution with slice-free charge-domain computation. Reconfigurable capacitor reuse unifies input DAC, multiplication, accumulation, shift-add, and SAR quantization, while differential inverse coding enables in-ADC partial-sum addition. Across CNN, Transformer, and LLM benchmarks, Charge-CIM reduces ADC energy by 91.7\%, improves energy efficiency by $2.71\times$, and improves throughput by $2.01\times$ over the state-of-the-art charge-domain baseline.



\bibliographystyle{IEEEtranS}
\bibliography{sample-base}

\end{document}